\documentclass[a4paper,12pt]{article}

\usepackage{amsmath}
\usepackage[psamsfonts]{amssymb}
\usepackage{rsfs}
\usepackage{bm}

\usepackage{cite}
\usepackage[dvips]{graphicx}
\usepackage{color}

\makeatletter
\@addtoreset{equation}{section}
\makeatother

\begin{document} 

\begin{titlepage}

\baselineskip 10pt
\hrule 
\vskip 5pt
\leftline{}
\leftline{Chiba Univ. Preprint
          \hfill   \small \hbox{\bf CHIBA-EP-169}}
\leftline{\hfill   \small \hbox{January 2008}}
\vskip 5pt
\baselineskip 14pt
\hrule 
\vskip 1.0cm
\centerline{\Large\bf 
} 
\vskip 0.3cm
\centerline{\Large\bf  
Wilson loop and magnetic monopole 
}
\vskip 0.3cm
\centerline{\Large\bf  
through a non-Abelian Stokes theorem
}
\vskip 0.3cm

\vskip 0.5cm

\centerline{{\bf 
Kei-Ichi Kondo,$^{\dagger,{1}}$  
}}  
\vskip 0.5cm
\centerline{\it
${}^{\dagger}$Department of Physics, Graduate School of Science, 
}
\centerline{\it
Chiba University, Chiba 263-8522, Japan
}
\vskip 1cm

\begin{abstract}
We show that the Wilson loop operator for $SU(N)$ Yang-Mills gauge connection is exactly rewritten in terms of conserved gauge-invariant magnetic and electric currents through a non-Abelian Stokes theorem of the Diakonov-Petrov type.
Here the magnetic current originates from the magnetic monopole derived in the gauge-invariant way from the pure   Yang--Mills theory even in the absence of the Higgs scalar field, in sharp contrast to the 't Hooft-Polyakov magnetic monopole in the Georgi-Glashow gauge-Higgs model. 
The resulting representation indicates that the Wilson loop operator in fundamental representations can be a probe for a single magnetic monopole irrespective of $N$ in $SU(N)$ Yang-Mills theory, against the conventional wisdom. 
Moreover, we show that the quantization condition for the   magnetic charge follows from the fact that the non-Abelian Stokes theorem does not depend on the surface chosen  for writing the surface integral. 
The obtained geometrical and topological representations of the Wilson loop operator have important implications to understanding quark confinement according to the dual superconductor picture. 
\end{abstract}

 
\vskip 0.5cm

\hrule  
\vskip 0.1cm
${}^1$ 
  E-mail:  {\tt kondok@faculty.chiba-u.jp}

\par 
\par\noindent


\vskip 0.5cm

\newpage
\pagenumbering{roman}
\tableofcontents




\end{titlepage}


\pagenumbering{arabic}

\baselineskip 14pt


\section{Introduction and main results}

The Wilson loop operator \cite{Wilson74} is a gauge-invariant observable of fundamental importance in Yang-Mills gauge theories \cite{YM54}.  It is defined from the holonomy of the gauge connection around a given loop. 
Especially, it is well-known that the area law of the Wilson loop average in Yang-Mills theory gives a criterion for quark confinement.

In the first half of this paper, we give a pedagogic and thorough derivation of a non-Abelian Stokes theorem for the Wilson loop operator of SU(N) Yang-Mills gauge connection.  
The non-Abelian Stokes theorem  (NAST) means that the non-Abelian Wilson loop operator defined for a closed loop $C$ is rewritten into a surface integral form over any surface bounding the loop $C$. 
In particular, we restrict our considerations to the Diakonov-Petrov (DP) type \cite{DP89} among various types of NAST.  
The DP version of NAST was originally derived in  \cite{DP89} for SU(2) case and later developed and extended to SU(N) case in \cite{DP89,DP96,DP00a,DP00b,FITZ00} and \cite{KondoIV,KT99,Kondo99Lattice99,HU99}. 
This is because the DP type can give us a very useful tool for understanding quark confinement based on the dual superconductor picture  \cite{dualsuper} which is a promising scenario of quark confinement.

In the latter half of this paper, some physical aspects obtained form the NAST will be demonstrated. 
The following are the main results of this paper obtained by extending the previous works \cite{KT99,Kondo99Lattice99} for SU(N) Yang-Mills gauge connection, although some of them have been known for the SU(2) case as will be mentioned in the relevant parts.  
We put our emphasis on the physical aspects rather than the mathematical rigor. 

1) 
The DP version of NAST includes no ordering procedures, which enables one to obtain an Abelian-like expression for the non-Abelian Wilson loop operator. 
It is derived as a  path-integral representation of the Wilson loop operator by using the coherent state for semi-simple Lie groups. 
The DP version of NAST is quite useful to consider the dual superconductivity as the electric-magnetic dual of ordinary superconductivity described by the  Maxwell-like Abelian gauge theory. 
See section~\ref{sec:NAST}.

2)
The contribution to the Wilson loop average is separated into two parts originating from the decomposition of the Yang-Mills potential $\mathscr{A}_\mu(x)$:
$
 \mathscr{A}_\mu(x) = \mathscr{V}_\mu(x) + \mathscr{X}_\mu(x)
$
in such a way that only the $\mathscr{V}_\mu$ part is responsible for quark confinement 
and that the remaining $\mathscr{X}_\mu$ part decouples.  
This is an gauge invariant understanding of the phenomenon  which was known so far as the infrared Abelian dominance \cite{tHooft81,EI82} in Yang-Mills theory under a specific gauge fixing called the maximal Abelian gauge \cite{KLSW87}.  In other words, this give a  gauge-invariant ``Abelian''  projection and ``Abelian dominance'' for the Wilson loop. 
In fact, the 100\% Abelian dominance for the Wilson loop is a direct consequence derived in the gauge-invariant way from  this construction \cite{Cho00,KS08}, although it was confirmed by numerical simulations on a lattice under the maximal Abelian gauge \cite{SY90,SNW94}. 
See section~\ref{sec:NASTr}.

3) 
The Wilson loop operator is explicitly rewritten in terms of conserved currents originating from the magnetic monopole which can be derived in the gauge-invariant way from the pure SU(N) Yang--Mills theory even in the absence of Higgs scalar field \cite{Cho80,FN98,Shabanov99,KMS06,KKMSSI05,IKKMSS06,SKKMSI07,SKKMSI07b}, in sharp contrast to the 't Hooft-Polyakov magnetic monopole in Georgi-Glashow gauge-Higgs model. 
This implies that the Wilson loop operator can be a probe for magnetic monopole whose condensation will cause the dual superconductivity in Yang-Mills theory.
See section~\ref{sec:magnetic-monopole}.

4) 
The quark confinement in the sense of area law of the Wilson loop average, if realized,  can be caused by a \textit{single} (\textit{non-Abelian} U($N-1$)) magnetic monopole \cite{WY07} irrespective of $N$ in SU($N$) Yang-Mills theory for $N \ge3$,  
against the conventional wisdom that $N-1$ magnetic monopoles associated with the maximal torus subgroup $U(1)^{N-1}$ are responsible for quark confinement.
This was conjectured in \cite{KT99,Kondo99Lattice99}. 
See section~\ref{sec:SU3}.

5) 
The quantization condition for the magnetic charge of the single magnetic monopole is derived from the fact that the non-Abelian Stokes theorem should not depend on the surface used to represent the surface integral. 
The quantization condition is the same as the Dirac type for SU(2), while it is similar to the Dirac type, but is different to that for $N \ge 3$. 
See section~\ref{sec:magnetic-charge}.

6) 
The Wilson loop operator has a geometrical and topological meaning related to solid angle seen at the location of the magnetic monopole in three-dimensional case, and the winding number between the magnetic monopole loop and the surface in four-dimensional case. 
This leads to a possibility that the area law can be derived by calculating the geometrical configurations of the relevant topological objects. 
See section~\ref{sec:magnetic-monopole}.

Section 3 is devoted to explaining the coherent state as a technical material for deriving the non-Abelian Stokes theorem. 
Some of more technical details are collected in Appendices. 

\section{Non-Abelian Stokes theorem}\label{sec:NAST}

The Wilson loop operator $W_{\rm C}[A]$ along the closed loop $C$ is defined as the trace of a path-ordered exponential of a gauge field $\mathscr{A}(x)$:
\begin{equation}
 W_C[A] := \mathcal{N}^{-1} {\rm tr} \left[ \mathscr{P} \exp \left\{ ig \oint_C \mathscr{A} \right\} \right] ,
\end{equation}
where $\mathscr{P}$ denotes the path ordering defined precisely later,
$\mathscr{A}$ is the Lie algebra valued connection one-form: 
\begin{equation}
 \mathscr{A}(x) :=  \mathscr{A}^A(x) T^A := \mathscr{A}_\mu^A(x) T^A dx^\mu
  ,
\end{equation}
and the normalization factor $\mathcal{N}$ is the dimension of the representation $R$, in which the Wilson loop is considered, i.e.,  
\begin{equation}
 \mathscr{N}:=d_R = {\rm dim}({\bf 1}_R) = {\rm tr}({\bf 1}_R)  .
\end{equation}

\begin{figure}[ptb]
\begin{center}
\includegraphics[height=1in]{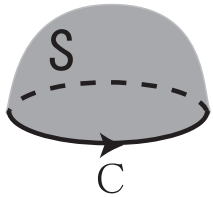}
\end{center} 
 \caption[]{
 Wilson loop $C$ and a surface $S$ bounded by $C$
}
 \label{fig:Wilson-loop}
\end{figure}

We wish to rewrite the non-Abelian Wilson loop operator defined by the line integral along a loop $C$ into a surface integral form over the surface $S$ having $C$ as the boundary: $\partial S=C$.  
See Fig.~\ref{fig:Wilson-loop}.
A curve (path) $L$ starting at $x_0$ and ending at $x$ is parameterized by a parameter $s$: $x_0=x(s_0)$ and $x=x(s)$.  Then we define the \textbf{parallel transporter} $W_L[A](s,s_0)$ by 
\begin{equation}
 W_L[A]_{ab}(s,s_0) 
=  \left[ \mathscr{P} \exp \left\{ ig \int_{L:x_0 \rightarrow x}  \mathscr{A}  \right\} \right]_{ab}  
=  \left[ \mathscr{P} \exp \left\{ ig \int_{s_0}^{s} ds \mathscr{A}(s) \right\} \right]_{ab} ,
\end{equation}
where a matrix element is specified by two indices $ab$  and we have defined
\begin{equation}
 \mathscr{A}(s) =  \mathscr{A}_\mu(x) dx^\mu/ds =  \mathscr{A}_\mu^A(x) T^A dx^\mu/ds .
\end{equation}

The Wilson loop operator $W_C[A]$ for a closed loop $C$ is obtained by taking the trace of $W_L[A]$ for a closed path $L=C$:
\begin{equation}
  W_C[A] = \mathcal{N}^{-1} {\rm tr} (W_C[A](s,s_0)) 
  = \mathcal{N}^{-1} \sum_{a=1}^{\mathcal{N}} W_C[A]_{aa}(s,s_0) .
\end{equation}

The parallel transporter $W_L[A]$ satisfies the differential equation of the Schr\"odinger type:
\begin{equation}
 i \frac{d}{ds}W_L[A](s,s_0) = -g \mathscr{A}(s) W_L[A](s,s_0) .
\end{equation}
Therefore, $W_L[A]$ is regarded as the time-evolution operator of a quantum mechanical system with the Hamiltonian
$H(s) = -g\mathscr{A}(s)$, if $s$ is identified with the time. 
This implies that it is possible to write path-integral representations of the parallel transporter and the Wilson loop operator according to the standard procedures:

\begin{enumerate}
\item
Partition the path $L$ into $N$ infinitesimal segments, 
\begin{equation}
 \mathscr{P} \exp \left[ ig \int_{s_0}^{s} ds \mathscr{A}(s) \right] 
=    \mathscr{P} \prod_{n=0}^{N-1} [1+ig \epsilon \mathscr{A}(s_n)]    ,
\end{equation}
where $\epsilon :=(s-s_0)/N$ and $s_n :=n\epsilon$. We set $s_0=0$ and  $s=s_N$. 

\item
Insert the complete (normal) set of states at each partition point, 
\begin{equation}
 {\bf 1} = \int \left| \xi_n, \Lambda \right> d\mu(\xi_n) \left< \xi_n, \Lambda \right| \quad (n=1, \cdots, N-1) ,
\end{equation}
where the state is normalized 
\begin{equation}
\left< \xi_n, \Lambda | \xi_n, \Lambda \right>=1 
 .
\end{equation}
\item
Take the limit $N \rightarrow \infty$ and $\epsilon \rightarrow 0$ appropriately such that $N\epsilon=s$ is fixed.

\item
Replace the trace of the operator $\mathscr{O}$ with 
\begin{equation}
 \mathcal{N}^{-1} {\rm tr}(\mathscr{O}) 
 = \int d\mu(\xi_N) \left< \xi_N, \Lambda \right| \mathscr{O}\left| \xi_N, \Lambda \right>  .
\end{equation}

\end{enumerate}

Putting aside the issue of what type of complete set is chosen, we obtain
\begin{align}
 &  \mathcal{N}^{-1}  {\rm tr} \left\{ \mathscr{P} \exp \left[  i g \int_0^s ds \mathscr{A}(s)\right] \right\}
\nonumber\\
  =& \lim_{N \rightarrow \infty, \epsilon \rightarrow 0} 
\int \cdots \int 
d\mu(\xi_{0}) \langle \xi_0,\Lambda |[1+i \epsilon g\mathscr{A}(s_0)] | \xi_1, \Lambda \rangle
\nonumber\\
&  \times
d\mu(\xi_{1}) \langle \xi_1,\Lambda |[1+i \epsilon g\mathscr{A}(s_1)] | \xi_2, \Lambda \rangle
\nonumber\\
& 
\cdots 
d\mu(\xi_{N-1})  \langle \xi_{N-1},\Lambda|[1+i \epsilon g\mathscr{A}(s_{N-1})] | \xi_{N}, \Lambda \rangle 
\nonumber\\
 =& \lim_{N \rightarrow \infty, \epsilon \rightarrow 0}
\prod_{n=1}^{N} \int d\mu(\xi_n) \prod_{n=0}^{N-1}
\langle \xi_{n}, \Lambda |[1+i \epsilon g\mathscr{A}(s_{n})] | \xi_{n+1}, \Lambda \rangle  
\nonumber\\
 =& \lim_{N \rightarrow \infty, \epsilon \rightarrow 0}
\prod_{n=1}^{N} \int d\mu(\xi_n) 
\prod_{n=0}^{N-1} [\langle \xi_{n}, \Lambda | \xi_{n+1}, \Lambda \rangle +i \epsilon g \langle \xi_{n}, \Lambda | \mathscr{A}(s_{n})] | \xi_{n+1}, \Lambda \rangle ]
 ,
\end{align}
where we have used $\xi_0=\xi_N$.

As a complete set to be inserted, we adopt the \textbf{coherent state}. As explained in the next section, the coherent state $\left| \xi_n, \Lambda \right>$ is constructed by operating a group element $\xi \in G$ to a reference state $\left| \Lambda \right>$:
\begin{equation}
 \left| \xi , \Lambda \right> = \xi \left| \Lambda \right> .
\end{equation}
Note that the coherent states are non-orthogonal:
\begin{equation}
\langle \xi^\prime ,\Lambda| \xi, \Lambda \rangle
\ne 0 .
\end{equation} 
For taking the limit $\epsilon \rightarrow 0$ in the final step to obtain the path-integral representation, it is sufficient to retain the $O(\epsilon)$ terms. 
Therefore, we find apart from $O(\epsilon^2)$ terms
\begin{align}
&  \epsilon \langle \xi_{n},\Lambda| \mathscr{A}(s_n) 
| \xi_{n+1}, \Lambda \rangle  
\nonumber\\
=&  \epsilon \langle \xi_{n},\Lambda| \mathscr{A}(s_n) 
| \xi_{n}, \Lambda \rangle + O(\epsilon^2) 
\nonumber\\
=&  \epsilon \langle \Lambda| \xi(s_{n})^\dagger \mathscr{A}(s_n) \xi(s_{n})  
| \Lambda \rangle + O(\epsilon^2)
 ,
\end{align}
and
\begin{align}
 \langle \xi_{n},\Lambda | \xi_{n+1}, \Lambda \rangle 
 =& \langle \xi(s_{n}),\Lambda | \xi(s_{n}), \Lambda \rangle 
+ \epsilon \langle \xi(s_{n}),\Lambda | \dot \xi(s_{n}), \Lambda
\rangle  +   O(\epsilon^2) 
\nonumber\\
 =& 1 + \epsilon \langle \xi(s_{n}),\Lambda | \dot \xi(s_{n}), \Lambda
\rangle  +   O(\epsilon^2)
 ,
\end{align}
where we have used the normalization condition,
$\langle \xi(s_{n}),\Lambda | \xi(s_{n}), \Lambda \rangle=1$.  Hence the dot denotes the differentiation with respect to $s$.
Therefore, we obtain
\begin{align}
 & \langle \xi_{n}, \Lambda | \xi_{n+1}, \Lambda \rangle +i \epsilon g \langle \xi_{n}, \Lambda | \mathscr{A}(s_{n})] | \xi_{n+1}, \Lambda \rangle
\nonumber\\
=& 1 + ig \epsilon \langle \Lambda| \xi(s_{n})^\dagger \mathscr{A}(s_n) \xi(s_{n})  
| \Lambda \rangle
+ \epsilon \langle \xi(s_{n}),\Lambda | \dot \xi(s_{n}), \Lambda
\rangle  +   O(\epsilon^2) 
\nonumber\\
=& \exp [ ig \epsilon \langle \Lambda| \xi(s_{n})^\dagger \mathscr{A}(s_n) \xi(s_{n})  
| \Lambda \rangle
+ \epsilon \langle \xi(s_{n}),\Lambda | \dot \xi(s_{n}), \Lambda
\rangle  +   O(\epsilon^2) ] 
 .
\end{align}
Thus we arrive at the expression: 
\begin{align}
 W_C[{\cal A}]
 = \int [d\mu(\xi)]_C
 \exp \left\{ i g \oint_{C} ds \langle \Lambda|
[ \xi(s)^\dagger \mathscr{A}(s) \xi(s)  
-   ig^{-1} \xi(s)^\dagger \dot \xi(s) ]| \Lambda
\rangle  \right\} 
 ,
\end{align} 
where 
$[d\mu(\xi)]_C$ is the
product measure of
$d\mu(\xi_n)$ along the loop $C$:
\begin{align}
  [d\mu(\xi)]_C := 
\lim_{N \rightarrow \infty, \epsilon \rightarrow 0}
\prod_{n=1}^{N}  d\mu(\xi_n)
=: \prod_{x \in C}d\mu(\xi(x))
 .
\end{align}

Another form for the path integral representation of the
Wilson loop operator reads
\begin{align}
 W_C[{\cal A}]  =  \int [d\mu(\xi)]_C \exp \left( 
i g \oint_C  \langle \Lambda | \left[ \xi(x)^\dagger \mathscr{A}(x) \xi(x)
-  ig^{-1} \xi(x)^\dagger d \xi(x)   \right]
|\Lambda \rangle \right) ,
\end{align}
where $d$ denotes the exterior derivative:
\begin{equation}
 d = ds \frac{d}{ds} = ds \frac{dx^\mu}{ds} \frac{\partial}{\partial x^\mu} =    dx^\mu \frac{\partial}{\partial x^\mu} .
\end{equation}
It is further rewritten into the form:
\begin{subequations}
\begin{align}
 W_C[\mathscr{A}] 
 = \int [d\mu(\xi)]_C \exp \left( 
i g \oint_C  \left[ m^A \mathscr{A}^A  
+  \omega \right] \right)  ,
\label{NAST0}
\end{align}
where we have introduced 
\begin{align}
 m^A(x)  :=&
 \langle \Lambda | \xi^\dagger(x) T^A \xi(x) |\Lambda \rangle ,
\label{mdef}
\\
  \omega(x)   :=& 
  - \langle \Lambda | ig^{-1} \xi^\dagger (x) d\xi(x) |\Lambda
\rangle .
\label{omegadef}
\end{align}
\end{subequations}

Now the argument of the exponential has been rewritten using Abelian quantities. 
Therefore, we can apply  the (usual) Stokes theorem,
\begin{equation}
 \oint_{C=\partial S} \omega = \int_S d \omega   
  ,
\end{equation}
 in the argument of the exponential.
Thus we obtain a version of  non-Abelian Stokes theorem (NAST):
\begin{subequations}
\begin{align}
 W_C[\mathscr{A}] 
 =& \int [d\mu(\xi)]_S \exp \left( 
ig \int_{S:\partial S=C} 
\left[ d(m^A \mathscr{A}^A) +  \Omega_K \right] \right) ,
\label{NAST}
\end{align}
where we have defined a curvature two-form:
\begin{equation}
 \Omega_K := d\omega
  ,
\end{equation}
and the product measure over the surface $S$:
\begin{align}
  [d\mu(\xi)]_S :=  
\prod_{x \in S}   d\mu(\xi(x)) .
\end{align}
\end{subequations}
For the representation to be meaningful, the field $\xi(x)$ must be continued over the surface $S$ inside the loop $C$, see \cite{DP00a} for detailed discussion on this issue.

\section{Coherent states}\label{sec:coherent-state}

\subsection{Coherent state and maximal stability group}

First, we construct the \textbf{coherent state}
$
  |\xi, \Lambda \rangle 
$
corresponding to the
coset representatives $\xi \in G/\tilde H$. 
We follow the method of Feng, Gilmore and Zhang \cite{FGZ90}.  As inputs, we prepare the following:
\begin{enumerate}
\item[(a)]
 The gauge group%
\footnote{
Note that any compact semi-simple Lie group is a direct product of
compact simple Lie group.  Therefore, it is sufficient to consider
the case of a compact simple Lie group.  In the following we assume
that
$G$ is a compact simple Lie group, i.e., a compact Lie group with no
closed connected invariant subgroup.
}
 $G$ has the Lie algebra
$\mathscr{G}$ with the generators $\{ T^A  \}$, which obey the
commutation relations 
\begin{equation}
 [T^A, T^B ] = i f^{AB}{}_C T^C ,
\end{equation} 
where the $f^{AB}{}_C$ are the structure constants of the Lie algebra. 
If the Lie algebra is semi-simple, it is more convenient to rewrite
the Lie algebra in terms of the \textbf{Cartan basis} 
$\{ H_j, E_\alpha, E_{-\alpha} \}$.  There are two types of basic 
operators in the Cartan basis, $H_j$ and $E_\alpha$.  The operators
$H_j$ may be taken as diagonal, while $E_\alpha$ are the
off-diagonal shift operators.  They obey the commutation relations 
\begin{align}
 [H_j, H_k] =& 0,
\\
[ H_j, E_\alpha ] =& \alpha_j E_\alpha,
\\
 [ E_\alpha, E_{-\alpha} ] =& \alpha^j H_j,
\\
 [ E_\alpha, E_\beta ] 
=& \begin{cases}
N_{\alpha;\beta} E_{\alpha+\beta}  &(\alpha+\beta \in R) \\ 
 0  &(\alpha+\beta \not\in R, \alpha+\beta \not= 0) 
   \end{cases}
    ,
\end{align}
where $R$ is the root system, i.e., a set of root vectors
$\{ \alpha_1, \cdots, \alpha_r \}$, with $r$ the rank of $G$.
\item[(b)]
  The Hilbert space $V^\Lambda$ is a carrier (the
representation space) of the unitary irreducible representation
$\Gamma^\Lambda$ of $G$.   
\item[(c)]  
We use a reference state 
$|\Lambda \rangle$ within the Hilbert space $V^\Lambda$,
which can be normalized to unity:
$
 \langle \Lambda |\Lambda \rangle = 1 .
$
\end{enumerate}
\par
We define the \textbf{maximal stability} subgroup (\textbf{ isotropy}
subgroup)
$\tilde H$ as a subgroup of
$G$ that consists of all the group elements $h$ that leave the
reference state $|\Lambda \rangle$ invariant up to a phase factor:
\begin{equation}
  h |\Lambda \rangle = |\Lambda \rangle e^{i\phi(h)}, \quad
 h \in \tilde H .
\end{equation}
Let $H$ be the \textbf{Cartan subgroup} of $G$, i.e., the maximal commutative
semi-simple subgroup in $G$,  and Let $\mathscr{H}$ be the \textbf{Cartan subalgebra}
in $\mathscr{G}$, i.e., the Lie algebra for the group $H$.
 The maximal stability subgroup $\tilde H$ includes the Cartan
subgroup $H=U(1)^r$, i.e., 
\begin{equation}
  H=U(1)^r \subset \tilde H .
\end{equation}

\par
For every element
$g\in G$, there is a unique decomposition of
$g$ into a product of two group elements, 
\begin{equation}
   g = \xi h  \in G, \quad \xi \in G/\tilde H, \quad h \in \tilde H .
\end{equation}
We can obtain a unique coset space $G/\tilde H$ for a given $|\Lambda \rangle$.
The action of arbitrary group element $g\in G$  on 
$|\Lambda \rangle$ is given by
\begin{equation}
  |g,\Lambda \rangle :=g |\Lambda \rangle = \xi h  |\Lambda \rangle
  = \xi  |\Lambda \rangle  e^{i\phi(h)}
= |\xi , \Lambda \rangle  e^{i\phi(h)}
 .
\end{equation}
The coherent state is constructed as  
\begin{equation}
 |\xi, \Lambda \rangle = \xi |\Lambda \rangle  
 .
\end{equation}
This definition of the coherent state is in one-to-one
correspondence with the coset space $G/\tilde H$ and the coherent states
preserve all the algebraic and topological properties of the coset
space $G/\tilde H$.
The phase factor is unimportant in the following discussion because
we consider the expectation value of operators $\mathscr{O}$ in the coherent state
\begin{equation}
\left< g, \Lambda |\mathscr{O} |g, \Lambda \right>
=\left< \xi, \Lambda |\mathscr{O} |\xi, \Lambda \right>
 .
\end{equation}

\par
If $\Gamma^\Lambda(\mathscr{G})$ is Hermitian, then $H_j^\dagger=H_j$ 
and $E_{\alpha}^\dagger = E_{-\alpha}$.  
Every group element $g \in G$ can be written as  the exponential of
a complex linear combination of diagonal operators $H_j$ and
off-diagonal shift operators $E_\alpha$.  
Let $|\Lambda \rangle$ be the highest-weight state, i.e., 
\begin{equation}
 H_j | \Lambda \rangle = \Lambda_j | \Lambda \rangle
 , \quad
 E_\alpha | \Lambda \rangle = 0 \ (\alpha \in R_+)
 , 
\end{equation}
where
$R_+ (R_-)$ is a subsystem of positive (negative) roots.
Then the coherent
state is given by \cite{FGZ90}
\begin{eqnarray}
  |\xi, \Lambda \rangle 
  = \xi |\Lambda \rangle
  = \exp \left[ 
  \sum_{\beta\in R_{-}} \left( \eta_\beta E_\beta - 
  \bar{\eta}_\beta E_{\beta}^{\dagger}\right) \right] |\Lambda \rangle,
\quad \eta_\beta \in \mathbb{C},
\label{cohrentdef}
\end{eqnarray}
such that
\begin{enumerate}
\item[(i)]
 $|\Lambda \rangle$  is
annihilated by all the (off-diagonal) shift-up operators $E_{\alpha}$ with
$\alpha \in R_+$, 
$
 E_{\alpha} |\Lambda \rangle = 0 \ (\alpha \in R_+) ;
$
\item[(ii)] 
$|\Lambda \rangle$  is
mapped into itself by all diagonal operators $H_j$,
$
 H_{j} |\Lambda \rangle = \Lambda_j |\Lambda \rangle ;
$
\item[(iii)] 
$|\Lambda \rangle$  is
annihilated by some shift-down operators $E_{\alpha}$ with
$\alpha \in R_-$, not by other $E_{\beta}$ with $\beta \in R_-$:
$
 E_{\alpha} |\Lambda \rangle = 0 \ ({\rm some~} \alpha \in R_-) ;
$
$
 E_{\beta} |\Lambda \rangle = |\Lambda+\beta \rangle 
 \ ({\rm some~} \beta \in R_-) ;
$
\end{enumerate}
and the sum $\sum_{\beta}$ is restricted to those shift operators
$E_{\beta}$ which obey (iii).
\par
The coherent state spans the entire space $V^\Lambda$.
By making use of the the group-invariant measure $d\mu(\xi)$ of
$G$ which is appropriately normalized, we obtain
\begin{eqnarray}
  \int |\xi, \Lambda \rangle d\mu(\xi) 
  \langle \xi,  \Lambda |
  = I 
 ,
 \label{resolution}
\end{eqnarray}
which shows that the coherent states are complete, but in fact over-complete.
The coherent states are normalized to unity: 
\begin{equation}
 \langle \xi, \Lambda | \xi, \Lambda \rangle = 1 .
\end{equation}
However, the coherent states are non-orthogonal: 
\begin{equation}
 \langle \xi' , \Lambda | \xi, \Lambda \rangle \not= 0 .
\end{equation}
The resolution of identity (\ref{resolution}) is very important to obtain the path integral formula of the Wilson loop operator given later.

\par
The coherent states 
$|\xi, \Lambda \rangle$
are in one-to-one correspondence with the coset representatives $\xi \in G/\tilde H$:
\begin{equation}
 |\xi, \Lambda \rangle \leftrightarrow G/\tilde H .
\end{equation}
In other  words, $|\xi, \Lambda \rangle$
and $\xi \in G/\tilde H$ are topologically equivalent.

Here it is quite important to remark that for G=SU(2) the stability subgroup agrees with the maximal torus group, i.e., 
\begin{equation}
 H=U(1)=\tilde H \ \text{for} \  G=SU(2)  .
\end{equation}
However, this is not necessarily the case for $G=SU(N)$  ($N \ge 3$). 
Therefore, this is an important point to be kept in mind for studying G=SU(3) case.

\subsection{$SU(2)$ coherent state}

\par
\begin{figure}[ptb]
\begin{center}
\includegraphics[height=1.5in]{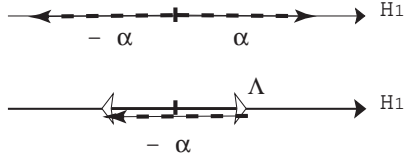}
\end{center}
\caption{Root diagram and weight diagram of the  fundamental representation of $SU(2)$ where  $\Lambda$ is the highest weight of the fundamental representation.}
 \label{fig:root_SU2}
\end{figure}

In the case of $SU(2)$, the rank is one, $r=1$.  
The maximal stability group $\tilde{H}$ agrees with the maximal torus subgroup $H=U(1)$ irrespective of the representation.
An irreducible  spin-$J$ representation $\{|J,M \rangle \}$ of $SU(2)$ is characterized by the highest spin $J$:
$J=\frac{1}{2}, 1, \frac{3}{2}, 2, \cdots$.
The root and weight diagrams are shown in Fig.\ref{fig:root_SU2}.


First, we study the representation of the $SU(2)$ coherent state by  real numbers. 
Let $|J,M \rangle$ be an eigenvector of both the quadratic Casimir invariant $\bm{J}^2$ and the diagonal generator $J_3$ (Cartan subalgebra) where $M$ labels the eigenvalue of $J_3$:  
\begin{align}
 \bm{J}^2 |J,M \rangle  =& J(J+1) |J,M \rangle  
 \quad (-J \le M \le J) ,
 \nonumber\\
 J_3 |J,M \rangle  =& M |J,M \rangle  .
\end{align}
Let $|\Lambda \rangle$ denote  the highest weight state $|\Lambda \rangle=|J,J \rangle$ of the spin-$J$ representation .

Spin coherent sates are a family of spin state
$\{ |{\bf n} \rangle \}$ which is obtained by applying
 the rotation matrix $R$ to the maximally polarized state $|\Lambda \rangle=|J,J \rangle$,
\begin{align}
 |{\bf n} \rangle := R(\alpha,\beta,\gamma) |J, J \rangle
 = e^{-iJ^3 \alpha} e^{-i J^2 \beta} e^{-iJ^3 \gamma} |J, J \rangle ,
 \label{rep}
\end{align}
where $J_A$ ($A=1,2,3$)  are three generators of su(2) and $(\alpha,\beta,\gamma)$ are Euler angles.
It turns out that the spin coherent state is characterized by the unit vector ${\bf n}$.  
It is shown that for $J=\frac12$,  
\begin{align}
 \langle {\bf n}(x) | J_A |{\bf n}(x) \rangle  =  J n_A(x) ,
\label{formula1}
\end{align}
where ${\bf n}(x)$ is a vector field of a unit length with three components:
\begin{align}
  {\bf n}(x)   = (n_1(x) ,n_2(x) ,n_3(x) ) 
  = (\sin \beta(x)  \cos \alpha(x) ,
\sin \beta(x)  \sin \alpha(x) , \cos \beta(x)  ).
\label{n-def}
\end{align}

We have the freedom to define $\gamma$ arbitrary.  This is a U(1) degrees of freedom.  
Therefore, the spin coherent states are in one-to-one correspondence with the (right) coset $SU(2)/U(1) \cong S^2$ where $U(1)$ is generated by $J_3$ (rotation about the third axis in the target space).

Moreover, it is known  that the coherent sates are not
orthogonal.  The overlap, i.e. the inner product of any two coherent
states is evaluated as
\begin{align}
 \langle {\bf n}  |{\bf n}' \rangle  =&
 \left( {1 + {\bf n} \cdot {\bf n}' \over 2} \right)^J
e^{-iJ\Phi({\bf n},{\bf n}')},
 \nonumber\\
 \Phi({\bf n},{\bf n}')  :=& 2 \arctan \left\{ 
 {\cos [{1 \over 2}(\beta+\beta')] \over
 \cos [{1 \over 2}(\beta-\beta')]}
 \tan \left( {\alpha - \alpha' \over 2} \right) \right\} +
\gamma - \gamma' ,
\label{formula2}
\end{align}
where $\gamma$ and $\gamma'$ correspond to the U(1) degrees of freedom mentioned above. 

\par
  The coherent states span the space of states of spin $J$.  The
measure of integration is defined by
\begin{align}
  d\mu({\bf n}) 
 := {2J+1 \over 4\pi}  \delta({\bf n}\cdot {\bf n}-1) d^3 {\bf n}
 = {2J+1 \over 4\pi} \sin \beta d\beta d\alpha .
 \label{measure}
\end{align}
This is a Haar measure of the coset $SU(2)/U(1)$,  
it is the area element on the two-sphere $S^2$.

The state $ |{\bf n} \rangle$ can be expanded in a complete basis of
the spin-$J$ irreducible representation 
$\{|J,M \rangle \}$.
The coefficients of the expansion are the representation matrix,
\begin{align}
  | {\bf n} \rangle = \sum_{M=-J}^{+J} |J,M \rangle 
 D^{(J)}_{MJ}({\bf n}) ,
 \label{expansion}
\end{align}
where $D^{(J)}_{MJ}({\bf n}) $ is the Wigner $D$-function. 
The resolution of unity is given by
\begin{align}
  \int d\mu({\bf n})
 |{\bf n} \rangle \langle {\bf n}| 
 = \sum_{M=-J}^{+J} |J,M \rangle \langle J,M | = I ,
 \label{formula3}
\end{align}
where $I$ is an identity operator.
Hence the coherent state $ |{\bf n} \rangle$ forms the complete set,
although it is not orthogonal. Thus the coherent states form an
overcomplete basis.
\par
In particular, for the fundamental representation  $J={1 \over 2}$, an element  in SU(2) is
written in terms of three local variables 
$(\alpha,\beta,\gamma)$ corresponding to the
Euler angles,
\begin{align}
 R(\alpha, \beta, \gamma)  
 =& e^{-i \alpha  \sigma_3/2} 
   e^{-i \beta \sigma_2/2} 
   e^{-i \gamma  \sigma_3/2}  
   \nonumber\\
  =& \begin{pmatrix}
 e^{-{i \over 2}(\alpha +\gamma )} \cos {\beta \over 2} &
 -e^{-{i \over 2}(\alpha -\gamma )} \sin {\beta \over 2} \cr
 e^{{i \over 2}(\alpha -\gamma )} \sin {\beta \over 2} &
 e^{{i \over 2}(\alpha +\gamma )} \cos {\beta  \over 2}     
 \end{pmatrix} 
 ,
\nonumber\\& 
\alpha \in [0,2\pi], \beta \in [0,\pi], 
 \gamma \in [0, 2\pi] ,
\label{Euler}
\end{align}
and (\ref{rep}) reads
\begin{equation}
 |{\bf n} \rangle =  R(\alpha, \beta, \gamma) 
 \begin{pmatrix} 
 1 \cr 0
 \end{pmatrix}
 = e^{-{i \over 2}\gamma } 
  \begin{pmatrix} 
 e^{-{i \over 2}\alpha } \cos {\beta \over 2} \cr 
  e^{{i \over 2}\alpha } \sin {\beta \over 2}  
  \end{pmatrix}  .
\end{equation}
Making use of the explicit representation, we can make sure that the formulae (\ref{formula1}), (\ref{n-def}), (\ref{formula2}), (\ref{measure}), (\ref{expansion}) and (\ref{formula3}) hold for $J=1/2$.

Next, we study the representation of the $SU(2)$ coherent state by a complex number. 
  The coherent state for $F_1:=SU(2)/U(1)$ is obtained as
\begin{eqnarray}
 |J; w \rangle = \xi(w) |J, -J \rangle 
 = e^{\zeta J_{+} - \bar \zeta J_{-}}  |J, -J \rangle 
 = {1 \over (1+|w|^2)^J}e^{ w J_{+}}  |J, -J \rangle ,
\end{eqnarray}
where $|J, -J \rangle :=|J, M=-J \rangle$ is the lowest
state of $|J, M \rangle$, 
\begin{equation}
  J_{+} = J_1+iJ_2 , \quad
  J_{-}=J_{+}^\dagger ,  \quad
 w = {\zeta \sin |\zeta| \over |\zeta| \cos |\zeta|} \in \mathbb{C}
 ,
\end{equation}
and $(1+|w|^2)^{-J}$ is a normalization factor that ensures 
$
 \langle J, w |J, w \rangle = 1 ,
$
which is obtained from the Baker-Campbell-Hausdorff (BCH) formulas \cite{Gilmore02}.
The invariant measure is given by
\begin{eqnarray}
 d\mu = {2J+1 \over 4\pi} {dwd\bar w \over (1+|w|^2)^2} .
\end{eqnarray}
For
$
J_A = {1 \over 2}\sigma^A \ (A=1,2,3)  
$
with Pauli matrices $\sigma^A$, we obtain
$
  J_{+} = 
\begin{pmatrix}
0 & 1 \cr 0 & 0
\end{pmatrix}
$ 
and
\begin{equation}
 \xi = e^{ w J_{+}}  
= 
\begin{pmatrix}
1 & w \cr 0 & 1  
\end{pmatrix}^T
\in F_1 = CP^1=SU(2)/U(1) \cong S^2
 .
\label{s2}
\end{equation}
\par

A representation of $O(3)$ vector $\mathbf{n}$ is given by
\begin{equation}
 n_A(x) =   \phi_a^*(x) \sigma_{ab}^A \phi_b(x) \quad (a,b=1,2)  
  ,
\end{equation}
which is equivalent to
\begin{equation}
 n_1 = 2 \Re(\phi^*_1   \phi_2), \quad
 n_2 = 2 \Im(\phi^*_1   \phi_2), \quad
 n_3 = |\phi_1|^2 - |\phi_2|^2 .
 \label{defn}
\end{equation}
The complex coordinate $w$ obtained by the stereographic projection from the north pole is identical to the inhomogeneous local coordinates of $CP^1$ when $\phi_2 \not=0$,
\begin{equation}
 w = w^{(1)} + i w^{(2)}
 = {n_1 + i n_2 \over 1-n_3}
 = {2 \phi_1 \phi^*_2 \over (|\phi_1|^2 + |\phi_2|^2)
 - (|\phi_1|^2 - |\phi_2|^2)}
 = {\phi_1 \over \phi_2} .
\end{equation}
The complex variable $w$ is a $CP^1=P^1(\mathbb{C})$ variable written as 
\begin{equation}
 w = e^{i\alpha} \cot \frac{\beta}{2}
 ,
\end{equation}
in terms of the polar coordinate $(\beta,\alpha)$ on $S^2$ or Euler angles. 
While the stereographic projection from the south pole leads to 
\begin{equation}
 w = {n_1 + i n_2 \over 1+n_3}
 = \left({\phi_2 \over \phi_1} \right)^*  
 = e^{i\alpha} \tan \frac{\beta}{2}
  ,
\end{equation}
if $\phi_1 \not=0$.
The variable $w$ is $U(1)$ rotation invariant.

Another representation of {\bf n} is obtained by using the parameterization 
(\ref{s2}) of the $F_1$ variable $\xi$:
\begin{equation}
 n_A =   \langle \Lambda | \xi(w)^\dagger \sigma^A \xi(w) | \Lambda \rangle 
 =  
 \begin{pmatrix}
 \phi^*_1 & 0 
 \end{pmatrix}
 \begin{pmatrix}
 1 & w^* \cr 0 & 1 
 \end{pmatrix}
 \sigma^A 
 \begin{pmatrix}
 1 & 0 \cr w & 1
 \end{pmatrix}
 \begin{pmatrix}
 \phi_1 \cr 0
 \end{pmatrix}
 .
\end{equation}
This leads to 
\begin{equation}
 n_1 =  |\phi_1|^2(w+w^*), \quad
 n_2 = -i |\phi_1|^2(w-w^*), \quad
 n_3 =  |\phi_1|^2(1-ww^*) .
\end{equation}
Indeed, this agrees with (\ref{defn}) if 
$w=({\phi_2 \over \phi_1})^*$.
The entire space of $F_1$ is covered by two charts,
\begin{equation}
 CP^{1} = U_1 \cup U_2, \quad
 U_a = \{ (\phi_1, \phi_2) \in CP^{1}; \phi_a \not= 0 \} .
\end{equation}
For more details, see  
Ref.\cite{KondoIV,KT99}.

\subsection{$SU(3)$ coherent state}

\begin{figure}[ptb]
\begin{center}
\includegraphics[height=1.5in]{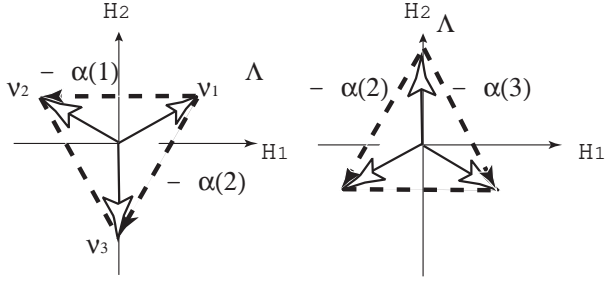}
\end{center}
\caption{The weight diagram and root vectors required to define the coherent state in the fundamental representations $[1,0]={\bf 3}$, $[0,1]={\bf 3}^*$  of $SU(3)$ where  $\vec
\Lambda = \vec h_1=\nu^1:=({1 \over 2},{1 \over 2\sqrt{3}})$ is the highest weight of the fundamental representation, and the other weights are 
 $\nu^2:=(-{1 \over 2},{1 \over 2\sqrt{3}})$ and
 $\nu^3:=(0,-{1 \over \sqrt{3}})$.}
 \label{fig:fundamental-weight}
\end{figure}

\begin{figure}
\begin{center}
\includegraphics[height=1.5in]{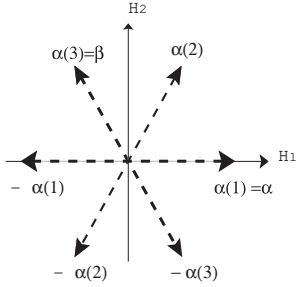}
\end{center} 
 \caption[]{The root diagram of $SU(3)$, where positive root vectors are given by 
$\alpha^{(1)}=(1,0)$,
$\alpha^{(2)}=({1 \over 2},{\sqrt{3} \over 2})$, and
$\alpha^{(3)}=({-1 \over 2},{\sqrt{3} \over 2})$.
The two simple roots are given by $\alpha^1:=
 \alpha=\alpha^{(1)},
$
and
$ \alpha^2:=\beta=\alpha^{(3)}$.}
 \label{fig:root}
\end{figure}

The rank of SU(3)  is two $r=2$ and every representation is specified by the Dynkin indices $[m,n]$.  The two-dimensional highest weight vector of the representation $[m,n]$ is given by 
\begin{equation}
 \bm{\Lambda} = m \bm{h_1} + n \bm{h_2}  
 ,
\end{equation}
using the highest weight $\bm{h_1}$ of a fundamental representation $[1,0]={\bf 3}$ and $\bm{h_2}$ of another  fundamental (conjugate) representation $[0,1]={\bf 3^*}$. 
The weight diagrams for fundamental representations ${\bf 3}$ and ${\bf 3^*}$ are given in Fig.~\ref{fig:fundamental-weight}. 
As the highest weight of ${\bf 3}$ we adopt the standard one: 
\begin{equation}
  \bm{h_1} 
= \left( \frac{1}{2}, \frac{1}{2\sqrt{3}} \right)
=: \nu_1  
 .
\end{equation}
As the highest weight of the conjugate representation ${\bf 3^*}$, on the other hand, we choose as in \cite{KT99} 
\footnote{
The following results hold also for other choices of $\bm{h_2}$, e.g., we can choose
\begin{equation}
  \bm{h_1} 
= \left( \frac{1}{2}, \frac{1}{2\sqrt{3}} \right)
=: \nu_1  
 , \quad
  \bm{h_2} 
= \left( \frac{1}{2}, \frac{-1}{2\sqrt{3}} \right) 
= - \nu_2 
 ,   \quad
  \bm{\Lambda} 
= \left( \frac{m+n}{2}, \frac{m-n}{2\sqrt{3}} \right) 
 .
\end{equation} 
}
\begin{equation}
  \bm{h_2} 
= \left( 0, \frac{1}{\sqrt{3}} \right) 
= - \nu_3
 .
\end{equation}
Then the highest weight of $[m.n]$ is given by 
\begin{equation}
 \bm{\Lambda} 
 = (\Lambda_1,\Lambda_2) 
=  \left({m \over 2}, {m+2n \over 2\sqrt{3}} \right) 
 .
\end{equation}
The generators  of SU(3) in the Cartan basis are written as
$\{ H_1, H_2, E_{\alpha}, E_{\beta}, E_{\alpha+\beta},$ $  E_{-\alpha}, E_{-\beta}, E_{-\alpha-\beta} \}$,
where $\alpha_1=\alpha$ and $\alpha_2=\beta$ are the two simple roots.
(See Fig.~\ref{fig:root} for the explicit choice.)

\begin{figure}
\begin{center}
\includegraphics[height=1.5in]{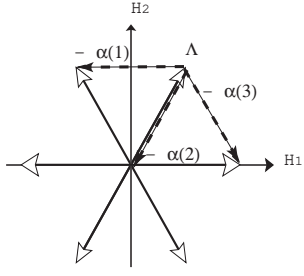}
\end{center} 
 \caption[]{The weight diagram and root vectors required to define
the coherent state in the adjoint representation $[1,1]={\bf 8}$ of
$SU(3)$, where  $\Lambda=({1 \over 2},{\sqrt{3} \over 2})$ is the highest weight of the adjoint
representation.}
 \label{fig:adjoint-weight}
\end{figure}

If $mn=0$, ($m=0$ or $n=0$), the maximal stability group
$\tilde H$ is given by $\tilde H=U(2)$ with generators 
$\{ H_1, H_2, E_\beta, E_{-\beta} \}$ (\textbf{minimal} case I, in which the coset $G/\tilde{H}$ is minimal). 
Such a degenerate case occurs when the highest-weight vector $\vec
\Lambda$ is orthogonal to some root vectors  
(see Fig.~\ref{fig:fundamental-weight}).
If $mn \not=0$ 
($m\not=0$ and $n\not=0$), $H$ is the maximal torus group 
$\tilde H=U(1) \times U(1)$  with generators 
$\{ H_1, H_2 \}$ (\textbf{maximal} case II, in which the coset $G/\tilde{H}$ is maximal). This is a non-degenerate case (see Fig.\ref{fig:adjoint-weight}).
Therefore, for the highest weight
$\Lambda$ in the minimal case (I), the coset $G/\tilde H$ is given by
\begin{equation}
  SU(3)/U(2)=SU(3)/(SU(2)\times U(1))=CP^2,
\end{equation}
whereas in the maximal case (II), 
\begin{equation}
  SU(3)/(U(1)\times U(1))=F_2 .
\end{equation}
Here, $CP^{n}$ is the complex projective space and $F_n$ is the flag space \cite{Perelomov87}.  
Therefore, \textit{the two fundamental representations of SU(3) belong to the minimal case (I), and hence the maximal stability group is $U(2)$, rather than the maximal torus group $U(1) \times U(1)$}.  

The implications of this fact to the mechanism of quark confinement is discussed in subsequent sections.

The coherent state for 
$F_2 = SU(3)/U(1)^2$ is given by
\begin{equation}
 | \xi, \Lambda \rangle = \xi(w) | \Lambda \rangle 
 := V^\dagger(w) | \Lambda \rangle ,
\end{equation}
with the highest- (lowest-) weight state $| \Lambda \rangle$, i.e.,
\begin{eqnarray}
 | \xi, \Lambda \rangle
 &=& \exp \left[\sum_{\alpha\in R_{+}}(\zeta_\alpha E_{-\alpha} -
\zeta^*_\alpha E_{-\alpha}^\dagger ) \right] | \Lambda \rangle
\nonumber \\
 &=&  e^{-{1 \over 2}K(w,w^*)}  
\exp \left[\sum_{\alpha\in R_{+}} \tau_\alpha E_{-\alpha} \right]  |
\Lambda \rangle ,
\label{cohe}
\end{eqnarray}
where
$e^{-{1 \over 2}K}$ is the normalization factor obtained from
the K\"ahler potential $K(w,w^*)$ (explained below):
\begin{eqnarray}
  K(w,w^*) 
  &:=& \ln [(\Delta_1(w,w^*))^m(\Delta_2(w,w^*))^n] ,
\nonumber\\
 \Delta_1(w,w^*) &:=& 1+|w_1|^2+|w_2|^2,
\nonumber\\
 \Delta_2(w,w^*) &:=& 1+|w_3|^2+|w_2-w_1 w_3|^2 . 
\label{Delta}
\end{eqnarray}
The coherent state 
$
 | \xi, \Lambda \rangle 
$
is normalized, so that
$
 \langle \xi, \Lambda | \xi, \Lambda \rangle = 1 
$.

It is shown \cite{KT99} that the inner product is given by
\begin{eqnarray}
 \langle \xi', \Lambda | \xi, \Lambda \rangle 
 =  e^{K(w, w^*{}')} e^{-{1\over2}[K(w',w^*{}')+K(w,w^*)]} ,
 \label{norm}
\end{eqnarray}
where 
\begin{equation}
  K(w, w^*{}') := 
 \ln [1+w^*_1{}' w_1+w^*_2{}'w_2]^m[1+w^*_3{}' w_3
 +(w^*_2{}'- w^*_1{}' w^*_3{}')(w_2-w_1w_3)]^n .
\end{equation}
Note that $K(w, w^*{}')$ reduces to the K\"ahler potential $K(w,w^*)$ when $w'=w$, in agreement with the normalization
$
 \langle \xi, \Lambda | \xi, \Lambda \rangle = 1 
$.

It follows from the general formula that the $SU(3)$ invariant measure is given (up to a constant factor) by
\begin{eqnarray}
 d\mu(\xi) = d\mu(w, w^*{})
 = D(m,n)[(\Delta_1)^m (\Delta_2)^n]^{-2} \prod_{\alpha=1}^{3}
dw_\alpha dw^*{}_\alpha , 
\end{eqnarray}
where $D(m,n)={1 \over 2}(m+1)(n+1)(m+n+2)$ is the dimension of the
representation. For the choice of shift-up $(E_{+i})$ or shift-down 
$(E_{-i})$ operators
\begin{equation}
  E_{\pm 1} := \frac{\lambda_1\pm i\lambda_2}{2\sqrt{2}}, \quad
  E_{\pm 2} := \frac{\lambda_4\pm i\lambda_5}{2\sqrt{2}}, \quad
  E_{\pm 3} := \frac{\lambda_6\pm i\lambda_7}{2\sqrt{2}}, 
\end{equation}
with the Gell-Mann matrices $\lambda_A$ $(A=1,\cdots,8)$,
we obtain
\begin{equation}
\exp \left[\sum_{i=1}^{3} \tau_{i} E_{-i} \right]  
 = 
 \begin{pmatrix}
 1 & w_1 & w_2 \cr 0 & 1 & w_3 \cr 0 & 0 & 1
 \end{pmatrix}^T
 \in F_2 = SU(3)/U(1)^2 ,
\end{equation}
where we have used the abbreviation $E_{\pm i}\equiv E_{\pm\alpha ^{(i)}}
\ (i=1,2,3)$.
These two sets of three complex variables are related as (see \cite{KT99})
\begin{equation}
 w_1 = \frac{\tau_1}{\sqrt{2}}, \quad w_2 = \frac{\tau_2}{\sqrt{2}} + \frac
 {\tau_1\tau_3}{4}, \quad w_3 = \frac{\tau_3}{\sqrt{2}} ,
\end{equation}
or conversely
\begin{equation}
 \tau_1 = \sqrt{2}w_1,\quad \tau_2 = \sqrt{2}\left(w_2 - \frac{w_1 w_3}{2}\right), 
\quad \tau_3 = \sqrt{2}w_3 .
\end{equation}

An element of $CP^2$ can be expressed using two complex variables, e.g., $w_1$ and $w_2$:
\begin{equation}
\exp \left[\sum_{i=1}^{2} \tau_{i} E_{-i} \right]  
 = 
 \begin{pmatrix}
 1 & w_1 & w_2 \cr 0 & 1 & 0  \cr 0 & 0 & 1
 \end{pmatrix}^T
 \in CP^2 = SU(3)/U(2)  ,
\end{equation}
The complex projective space $CP^2$ is covered by three complex planes
$\mathbb{C}$ through holomorphic maps \cite{Ueno95} (see e.g., \cite{KT99}).

The parameterization of $SU(3)$ in terms of eight angles is  possible  also in $SU(3)$, just as $SU(2)$ is parameterized by three Euler angles (see Ref.\cite{FiniteRotation}).

The SU(N) case can be treated in the similar way, see Ref.\cite{KT99} for details.

\section{Color fields construction from the coherent state}\label{sec:SU3}

Here we introduce $\mathcal{H}$ given by 
\begin{equation}
 \mathcal{H} := 
\bm{\Lambda} \cdot \bm{H}  = \sum_{j=1}^{r} \Lambda_j H_j 
 ,
\end{equation}
where $H_j$ ($j=1,, \cdots, r$) are the generators from the Cartan subalgebra ($r=N-1$ is the rank of the gauge group $G=SU(N)$) and 
$r$-dimensional vector $\Lambda_j$ ($j=1,, \cdots, r$) is the highest weight of the representation in which the Wilson loop is considered.

For fundamental representations,  we can write 
\begin{equation}
  \bm{e}_{ff} := \frac{1}{{\rm tr}(\bm{1})} \bm{1} + 2 \mathcal{H} 
 , \quad
  \mathcal{H}  = \frac12 \left( \bm{e}_{ff} - \frac{1}{{\rm tr}(\bm{1})} \bm{1}  \right) 
   ,
\end{equation}
where we introduce a matrix $\bm{e}_{ff}$ which has only one  non-vanishing $ff$ diagonal element with the value one: 
\begin{equation}
(\bm{e}_{ff})_{ab}=\delta_{fa}\delta_{fb}  \quad
 \text{(no sum over $f$)}
 .
\end{equation}
A highest-weight state $\left|  \bm{\Lambda} \right> $ is the common (normalized) eigenvector of $H_1, H_2, \cdots, H_r$ with the eigenvalue $\Lambda_1, \Lambda_2, \cdots, \Lambda_r$, i.e.,
\begin{equation}
H_j \left|  \bm{\Lambda} \right> 
=\Lambda_j \left|  \bm{\Lambda} \right> 
 \ ( j=1, \cdots, r=N-1 ) 
   .
   \label{eigen}
\end{equation}

For SU(2), every representation is specified by an half integer $J$:
\begin{equation}
 \Lambda_1 =J = \frac12 , 1, \frac32, 2, \cdots 
  , \quad
  H_1=\frac{\sigma_3}{2}
   ,
\end{equation}  
and the fundamental representation $J=\frac12$ of SU(2) leads to 
\begin{equation}
 \mathcal{H}  
= \frac12 \left( \bm{e}_{11} - \frac{1}{2} \bm{1}  \right) 
 = \frac{1}{2} {\rm diag} \left( \frac{1}{2}, \frac{-1}{2} \right) 
= \frac12 \frac{\sigma_3}{2}  
  .
\end{equation}

For SU(3), the explicit form of $\mathcal{H}$  for the fundamental representations reads using the diagonal set of the Gell-Mann matrices $\lambda_3$ and $\lambda_8$:
\begin{align}
 \mathcal{H} 
=& \frac{1}{2} (\Lambda_1 \lambda_3 + \Lambda_2 \lambda_8)
\nonumber\\
=& \frac{1}{2}
 \begin{pmatrix} 
  \frac{2m+n}{3} & 0 & 0 \cr
  0 & \frac{-m+n}{3} & 0 \cr
  0 & 0 & \frac{-m-2n}{3}  \cr
 \end{pmatrix} 
= \frac{1}{2} {\rm diag} \left( \frac{2m+n}{3}, \frac{-m+n}{3} , \frac{-m-2n}{3} \right)
 .
\end{align}
We enumerate all fundamental representations $\textbf{3}$:
\footnote{For another choice of $h_2=-\nu_2$, the same results are obtained if the following replacement is performed.
[-1,1] $\rightarrow$ [0,-1], 
[0,-1] $\rightarrow$ [-1,1],
[0,1] $\rightarrow$ [1,-1],
[1,-1] $\rightarrow$ [0,1].
}
\begin{subequations}
\begin{enumerate}
\item[[1,0]]:
\begin{equation}
 \bm{\Lambda} 
= \left( \frac{1}{2}, \frac{1}{2\sqrt{3}} \right) := \nu_1 
 , \quad
  \mathcal{H} = \frac{1}{2} {\rm diag} \left( \frac{2}{3}, \frac{-1}{3} , \frac{-1}{3} \right) 
   ,
   \label{fr1}
\end{equation}

\item[[-1,1]]:
\begin{equation}
 \bm{\Lambda} 
= \left( \frac{-1}{2}, \frac{1}{2\sqrt{3}} \right) := \nu_2 
 , \quad
  \mathcal{H} = \frac{1}{2} {\rm diag} \left( \frac{-1}{3}, \frac{2}{3} , \frac{-1}{3} \right)
   ,
   \label{fr2}
\end{equation}

\item[[0,-1]]:
\begin{equation}
 \bm{\Lambda} 
= \left( 0, \frac{-1}{\sqrt{3}} \right) := \nu_3 
 , \quad
  \mathcal{H} = \frac{1}{2} {\rm diag} \left( \frac{-1}{3}, \frac{-1}{3} , \frac{2}{3} \right)
  = \frac{-1}{\sqrt{3}} \frac{\lambda_8}{2} 
   ,
   \label{fr3}
\end{equation}

\end{enumerate}
\end{subequations}
and their conjugates $\textbf{3$^*$}$
\begin{subequations}
\begin{enumerate}
\item[[0,1]]:
\begin{equation}
 \bm{\Lambda} 
= \left( 0, \frac{1}{\sqrt{3}} \right) = -\nu_3 
 , \quad
  \mathcal{H} = \frac{1}{2} {\rm diag} \left( \frac{-1}{3}, \frac{-1}{3} , \frac{2}{3} \right)
  = \frac{-1}{\sqrt{3}} \frac{\lambda_8}{2} 
   ,
\end{equation}

\item[[1,-1]]:
\begin{equation}
 \bm{\Lambda} 
= \left( \frac{1}{2}, \frac{-1}{2\sqrt{3}} \right) = -\nu_2 
 , \quad
  \mathcal{H} = \frac{1}{2} {\rm diag} \left( \frac{-1}{3}, \frac{2}{3} , \frac{-1}{3} \right)
   ,
\end{equation}

\item[[-1,0]]:
\begin{equation}
 \bm{\Lambda} 
= \left( \frac{-1}{2}, \frac{-1}{2\sqrt{3}} \right) = -\nu_1 
 , \quad
  \mathcal{H} = \frac{1}{2} {\rm diag} \left( \frac{2}{3}, \frac{-1}{3} , \frac{-1}{3} \right)
   .
\end{equation}

\end{enumerate}
\end{subequations}

For three fundamental representations (\ref{fr1}), (\ref{fr2}) and (\ref{fr3}), the eigenvectors (\ref{eigen}) are found to be
\begin{align}
\left|  \bm{\Lambda} \right> 
=(1,0,0)^T , 
\quad
\left|  \bm{\Lambda} \right> 
=(0,1,0)^T ,
\quad \& \quad 
\left|  \bm{\Lambda} \right> 
=(0,0,1)^T
 ,
\end{align}
respectively.

Now we check that the coherent state indeed satisfies the desired properties.
In what follows, we restrict our consideration to the fundamental representations of SU(N), $N \ge 3$. 
For SU(2), any representation  will be discussed separately. 

As a reference state or a highest-weight state for a  fundamental representation of SU(N), we choose, e.g., $f=N$ 
\begin{equation}
\left|  \bm{\Lambda} \right> 
= (0, \cdots,0,  1)^T
= \begin{pmatrix}
  0 \cr
  \vdots \cr
  0 \cr
  1 
 \end{pmatrix}
 \quad \text{or} \quad 
\left< \bm{\Lambda} \right| = (0, \cdots, 0, 1)  
 ,
\end{equation}
which yields a projection operator:
\begin{equation}
\left|  \Lambda \right> \left< \Lambda \right| 
= e_{NN} 
= \begin{pmatrix}
  0 &  \cdots & 0 &0 \cr
  \vdots & \vdots & \vdots & \vdots \cr
  0  & \cdots & 0 & 0\cr
  0  & \cdots & 0 & 1
 \end{pmatrix}
 .
\end{equation}
First, we show the completeness
\footnote{
The following relationships hold even if we replace $\xi \in G/\tilde{H}$ by a general element $g \in G$, since they hold on a reference state $\left|  \bm{\Lambda} \right>$.
} 
\begin{equation}
 \int \left| \xi , \Lambda \right> d\mu(\xi ) \left< \xi , \Lambda \right|  = {\bf 1}/d_R
  .
\end{equation}
The $ab$ element of the left-hand side reads 
\begin{align}
 \left( \int \left| \xi , \Lambda \right> d\mu(\xi ) \left< \xi , \Lambda \right| \right)_{ab}
 =&  \int  d\mu(\xi )  \left(  \xi  \left|  \Lambda \right> \left< \Lambda \right| \xi^\dagger    \right)_{ab} 
 \nonumber\\
 =&  \int  d\mu(\xi )  \left(  \xi  \bm{e}_{ff} \xi^\dagger    \right)_{ab} 
 \nonumber\\
 =&  \int  d\mu(\xi )  (\xi)_{ac}  (\bm{e}_{ff})_{cd}  (\xi^\dagger)_{db} 
 \nonumber\\
 =&  \frac{1}{d_R} \delta_{ab} \delta_{cd} (\bm{e}_{ff})_{cd} 
 \nonumber\\
 =&  \frac{1}{d_R} \delta_{ab}  
  , 
  \label{comp}
\end{align}
where we have used the integration formula for the Haar measure $dU=d\mu(G)$ of SU(N):
\begin{equation}
 \int  dU \  U_{ac}(R)   U^\dagger_{db}(R') 
 = \frac{1}{d_R} \delta_{ab} \delta_{cd} \delta_{RR'}
 .
\end{equation}
Second, we show 
\begin{equation}
 \int d\mu(\xi ) \left< \xi , \Lambda \right| \mathscr{O}\left| \xi , \Lambda \right> 
 =  {\rm tr}(\mathscr{O})/{\rm tr}({\bf 1}) 
 .
\end{equation}
In fact, we have
\begin{align}
   \int d\mu(\xi ) \left< \xi , \Lambda \right| \mathscr{O}\left| \xi , \Lambda \right> 
 =& \int d\mu(\xi ) \left<  \Lambda \right| \xi^\dagger  \mathscr{O}\xi \left|  \Lambda \right> 
\nonumber\\
 =& \int d\mu(\xi ) (\xi^\dagger  \mathscr{O}\xi)_{ff}
\nonumber\\
 =& \int d\mu(\xi ) {\rm tr}(\xi^\dagger  \mathscr{O}\xi \bm{e}_{ff})
 \nonumber\\
 =& \int d\mu(\xi ) {\rm tr}( \mathscr{O}\xi \bm{e}_{ff}\xi^\dagger  )
 \nonumber\\
 =&  {\rm tr}( \mathscr{O} \int d\mu(\xi )\xi \bm{e}_{ff}\xi^\dagger  )
 \nonumber\\
 =&  {\rm tr}( \mathscr{O} \frac{1}{d_R} {\bf 1}  )
 = \frac{{\rm tr}(\mathscr{O})}{d_R} 
 ,
\end{align}
where we have used the previous result (\ref{comp}). 

Third, the normalization condition is trivial:
\begin{equation}
\left< \xi , \Lambda | \xi , \Lambda \right>
= \left<   \Lambda | \xi^\dagger \xi | \Lambda \right>
= \left<   \Lambda |   \Lambda \right>
=1 
 .
\end{equation}

Finally, an important relationship for the matrix element is derived. 
By using the coherent state for fundamental representations of SU(N), the  matrix element of any Lie algebra valued operator $ \mathscr{O}$  in the coherent state is cast into the form of the trace:
\begin{subequations}
\begin{align}
 \langle \xi,\Lambda| \mathscr{O}    | \tilde\xi, \Lambda \rangle 
=&  \langle \Lambda| \xi^\dagger  \mathscr{O}  \tilde\xi |  \Lambda \rangle 
\nonumber\\
=& (\xi^\dagger  \mathscr{O}  \tilde\xi )_{ff} \ (\text{no sum over $f$})
\nonumber\\
=& {\rm tr}[ \xi^\dagger \mathscr{O}  \tilde\xi  \bm{e}_{ff}    ]
\\
=& {\rm tr}[ \mathscr{O}  \tilde\xi  \bm{e}_{ff} \xi^\dagger   ]
 .
\end{align}
\end{subequations}
This is further rewritten as
\begin{subequations}
\begin{align}
  \langle \xi,\Lambda| \mathscr{O}    | \tilde\xi, \Lambda \rangle 
=& {\rm tr}[ \mathscr{O}  \tilde\xi  \bm{e} \xi^\dagger   ]
\nonumber\\
=& {\rm tr} \left[ \xi^\dagger \mathscr{O} \tilde\xi  \left( \frac{1}{{\rm tr}(\bm{1})} \bm{1}   + 2 \mathcal{H} \right)   \right]
\\
=& {\rm tr} \left[ \mathscr{O}   \left( \frac{1}{{\rm tr}(\bm{1})} \tilde\xi \xi^\dagger  + 2 \tilde\xi \mathcal{H} \xi^\dagger   \right)   \right]
 .
 \label{matrix-element}
\end{align}
\end{subequations}
The diagonal matrix element is cast into \cite{FGZ90,Perelomov87} 
\begin{align} 
  \langle \xi(x),\Lambda|\mathscr{O}(x)  | \xi(x), \Lambda \rangle 
=   {\rm tr} \left\{ \mathscr{O}(x)  \left[ \frac{1}{{\rm tr}(\bm{1})}\bm{1} + 2 \bm{m}(x) \right]   \right\} 
 .
\end{align}
where we have introduced a new field $\bm{m}(x)$ having its value in the Lie algebra $\mathscr{G}=su(N)$ by 
\footnote{
The  $\bm{m}(x)$ field can be normalized by multiplying a factor $\sqrt{\frac{2N}{N-1}}$, since 
\begin{equation}
2 {\rm tr} [\bm{m}(x) \bm{m}(x)]
= 2 {\rm tr}(  \mathcal{H} \mathcal{H}  )
= 2 \Lambda_j  \Lambda_k {\rm tr}( H_j H_k )
=   \Lambda_j^2 
= \frac{N-1}{2N} 
 .
\end{equation}
}
\begin{equation}
  \bm{m}(x)  
 := \xi(x) \mathcal{H} \xi(x)^\dagger 
=  \sum_{j=1}^{r}  \Lambda_j  \xi(x)H_j  \xi(x)^\dagger 
    .
\end{equation}
We can introduce the normalized color field $\bm{n}(x)$ by
\begin{equation}
  \bm{n}(x)  
  := \sqrt{\frac{2N}{N-1}} \bm{m}(x) 
  = \sqrt{\frac{2N}{N-1}}  \xi(x) \mathcal{H} \xi(x)^\dagger 
=  \sqrt{\frac{2N}{N-1}}  \sum_{j=1}^{r}  \Lambda_j  \xi(x)H_j  \xi(x)^\dagger 
\label{unit-n}
    .
\end{equation}
In particular, we have
\begin{align} 
  \langle \Lambda|\mathscr{O}(x)  |   \Lambda \rangle 
=   {\rm tr} \left\{  \left[ \frac{1}{{\rm tr}(\bm{1})}\bm{1} + 2 \mathcal{H} \right]  \mathscr{O}(x)  \right\} 
 .
\end{align}
Moreover, the traceless $\mathscr{O}(x)$ obeys more  simple relations:
\begin{align}  
\langle \xi(x),\Lambda|\mathscr{O}(x)  | \xi(x), \Lambda \rangle 
=&   2{\rm tr} \left\{ \bm{m}(x) \mathscr{O}(x) \right\}
 ,
\nonumber\\
\langle  \Lambda|\mathscr{O}(x)  |   \Lambda \rangle 
=&   2{\rm tr} \left\{ \mathcal{H} \mathscr{O}(x) \right\}
 .
\label{aaa}
\end{align}
Note that the $\bm{m}$ field defined from the coset element $\xi \in G/\tilde{H}$ is the same as that defined from the original group $g \in G$:
\begin{equation}
  \bm{m}(x)  
  =  g(x) \mathcal{H} g(x)^\dagger
  . 
\end{equation}
This is because
\begin{equation}
 g(x) \mathcal{H} g(x)^\dagger
  = \xi(x) h(x) \mathcal{H} h(x)^\dagger \xi(x)^\dagger
  = \xi(x)  \mathcal{H}  \xi(x)^\dagger
  , 
\end{equation}
which follows from a fact:
\begin{equation}
 h(x) \mathcal{H} h(x)^{-1}= \mathcal{H} \ 
\Longleftrightarrow 
 [h(x),\mathcal{H}]=0
  , 
\end{equation}

By introducing the  color fields $\bm{n}_j$ defined by
\begin{equation}
  \bm{n}_j(x) :=  g(x)  H_j  g^{\dagger}(x) 
    ,
\end{equation}
 the  $\bm{m}$ field is written as a linear combination
\begin{equation}
  \bm{m}(x)  
  =  g(x) \mathcal{H} g(x)^\dagger
  = \sum_{j=1}^{r}  \Lambda_j \bm{n}_j(x) 
 . 
   \label{m-2}
\end{equation}

For SU(2), every representation is specified by a half integer $J$ and the color field is unique, 
\begin{equation}
  \bm{n}(x)=\bm{n}_1(x) 
    ,
\end{equation}
and
\begin{equation}
  \bm{m}(x)   
=   \Lambda_1 \bm{n}(x) 
=  \pm J \bm{n}(x) 
=  \pm J \xi(x) \frac{\sigma_3}{2}  \xi(x)^\dagger 
    .
    \label{m-SU2}
\end{equation}
For $J=\frac12$, 
Using (\ref{formula1}), we obtain
\begin{align}
     \mathscr{A}^A  n^A 
 =   {\rm tr}[\sigma_3 G^\dagger \mathscr{A} G] ,
\end{align}
where $\sigma_3$ is the third Pauli matrix.
Indeed, we have 
\begin{align}
\frac12 {\rm tr}[\sigma_3 G^\dagger \mathscr{A} G]
=& \frac12 [(G^\dagger \mathscr{A} G)_{11}-(G^\dagger \mathscr{A} G)_{22}]
=  (G^\dagger \mathscr{A} G)_{11}
\nonumber\\
=&  \begin{pmatrix} 
 1 & 0 \cr 
 \end{pmatrix}
   [G^\dagger \mathscr{A} G] 
 \begin{pmatrix} 
 1 \cr 0
 \end{pmatrix} 
\nonumber\\
=& \langle {\bf n} |\mathscr{A}^A T^A |{\bf n} \rangle 
=  \langle {\bf n} |T^A |{\bf n} \rangle \mathscr{A}^A  
=  \frac12 n^A  \mathscr{A}^A ,
\end{align}
where  we have used the fact that the matrix $(G^\dagger \mathscr{A} G)$ is traceless, i.e., 
$(G^\dagger \mathscr{A} G)_{11}+(G^\dagger \mathscr{A} G)_{22}=0$, since 
$
{\rm tr}[G^\dagger \mathscr{A} G]={\rm tr}[\mathscr{A} G G^\dagger]
={\rm tr}[\mathscr{A}]
=\mathscr{A}^A {\rm tr}[T^A]=0
$.
In other words, ${\bf n}$ has the adjoint orbit representation: 
\begin{align}
 n^A(x) = {\rm tr}(\sigma_3 G^\dagger(x)T^A G(x))  ,
 \quad 
 n^A(x) T^A = G(x) T^3 G^\dagger(x)  . 
 \label{aorep}
\end{align} 
Using (\ref{Euler}), we can see that
the unit vector ${\bf n}(x)$ defined by (\ref{aorep}) is indeed equal to 
(\ref{n-def}).
Indeed, the adjoint orbit representation leads to the consistent result: 
\begin{align}
{\rm tr}(\sigma_3 G^\dagger \mathscr{A} G ) 
=& {\rm tr}(G \sigma_3 G^\dagger \mathscr{A} ) 
= 2 {\rm tr}(n^A T^A    \mathscr{A}^B T^B ) 
\nonumber\\
=& 2 n^A  \mathscr{A}^B  {\rm tr}(T^A    T^B ) 
= n^A  \mathscr{A}^B \delta^{AB} = n^A  \mathscr{A}^A .
\end{align}

For SU(3), the  $\bm{m}$ field is a linear combination of two color fields:
\begin{equation}
 \bm{n}_1(x)
  =  g(x)  \frac{\lambda_3}{2}  g^{\dagger}(x) 
  , \quad 
  \bm{n}_2(x)
  =  g(x)  \frac{\lambda_8}{2}  g^{\dagger}(x) 
   .
\end{equation}
The   $\bm{m}$ field reads
 for $[1,0]$ and $[-1,0]$, 
\begin{subequations}
\begin{equation}
  \bm{m}(x)  
  =  \pm   \frac{1}{2} \left[  \bm{n}_1(x)  +   \frac{1}{\sqrt{3}} \bm{n}_2(x) \right]
   =  \pm   \frac{1}{\sqrt{3}} \left[  \frac{\sqrt{3}}{2}  \bm{n}_1(x)  +   \frac{1}{2} \bm{n}_2(x) \right]
 , 
 \label{m1}
\end{equation}
for  $[0,-1]$ and $[0,1]$, 
\begin{equation}
  \bm{m}(x)  
  =  \pm  \frac{1}{2} \left[ - \bm{n}_1(x)  +   \frac{1}{\sqrt{3}} \bm{n}_2(x) \right] 
   =  \pm   \frac{1}{\sqrt{3}} \left[  -\frac{\sqrt{3}}{2}  \bm{n}_1(x)  +   \frac{1}{2} \bm{n}_2(x) \right]
  .
  \label{m2}
\end{equation}
In particular, for $[-1,1]$ and $[1,-1]$, $\Lambda_1=0$ and hence the    $\bm{m}$ field   is written using only $\bm{n}_2(x)$:
\begin{equation}
  \bm{m}(x)  
  =   \pm  \frac{1}{2} \left[  \frac{-2}{\sqrt{3}} \bm{n}_2(x) \right]
   =   \pm   \frac{1}{\sqrt{3}}  \left[   - \bm{n}_2(x) \right]
   \label{m3}
  .
\end{equation}
\end{subequations}

For SU(N), we can introduce $N-1$ color fields $\bm{n}_j(x)$ ($j=1, \cdots, N-1$) corresponding to the degrees of freedom of the maximal torus group $U(1)^{N-1}$ of SU(N).  This is just the way adopted in the conventional approach.  
However, this is not necessarily effective to see the physics extractable  from the Wilson loop.
This is because only the specific combination $\bm{m}(x)$ of the color fields $\bm{n}_j(x)$ has a physical meaning as shown in the above and this nice property of $\bm{m}(x)$ will be lost once $\bm{m}(x)$ is separated into the respective color field, except for the SU(2) case in which  $\bm{m}(x)$ agrees with the only one color field $\bm{n}(x)$ of SU(2). 
In view of this, only the last color field $\bm{n}_{N-1}(x)$ is enough for investigating quark confinement through the Wilson loop in fundamental representations of SU(N).   
\footnote{
This is called the minimal option proposed in \cite{KSM08}. Indeed, the above combinations (\ref{m1}), (\ref{m2}) and (\ref{m3}) correspond to 6 minimal cases discussed in \cite{KSM08}.
A unit vector $\bm{n}(x)$ introduced in (\ref{unit-n}) and ref.\cite{KSM08} is related to $\bm{m}$ as
$\sqrt{3} \bm{m}(x)=\bm{n}(x)=(\cos \vartheta(x)) \bm{n}_1(x)+(\sin \vartheta(x)) \bm{n}_2(x)$
where $\vartheta(x)$ denotes the angle of a weight vector in the weight diagram measured anticlockwise from the $H_1$ axis. 
Here 
(\ref{m1}), (\ref{m2}) and (\ref{m3}) correspond to $\vartheta(x)=\frac{1}{6}\pi (\frac{7}{6}\pi)$ $\frac{5}{6}\pi (\frac{11}{6}\pi)$, and $\frac{3}{2}\pi (\frac{1}{2}\pi)$ respectively. 
}

Note that  
\begin{equation}
  \bm{m}(x) = m^A(x) T^A , \quad 
  m^A(x) = 2 {\rm tr}(\bm{m}(x) T^A)
 = 2 {\rm tr}(g(x) \mathcal{H}  g^{\dagger}(x)T^A)
   ,
\end{equation}
for the normalization 
${\rm tr}(T^A T^B)=\frac12 \delta^{AB}$. 
For three fundamental representations (\ref{fr1}), (\ref{fr2}) and (\ref{fr3}) of SU(3), $m^A(x)$ is equal to the first, second and third diagonal elements of $g(x)T^A g^{\dagger}(x)$ respectively: 
\begin{equation}
  m^A(x) =  (g^{\dagger}(x)T^A g(x))_{ff} \ (f=1,2,3)
   .
\end{equation}
This is checked easily, e.g., for $[1,0]$,
\begin{align}
m^A(x) =& 2 {\rm tr}(\mathcal{H}  g^{\dagger}(x)T^A g(x) )
\nonumber\\
=& \frac23 (g^{\dagger}(x)T^A g(x))_{11} - \frac13 (g^{\dagger}(x)T^A g(x))_{22} - \frac13 (g^{\dagger}(x)T^A g(x))_{33} 
\nonumber\\
=&  (g^{\dagger}(x)T^A g(x))_{11}
 ,
\end{align}
where we have used a fact that $g^{\dagger}(x)T^A g(x)$ is traceless. 
Therefore, we have 
\begin{equation}
  m^A(x)  
 =  \left< \bm{\Lambda} |  g^{\dagger}(x)T^A g(x) |\bm{\Lambda} \right>
 =  \left< g(x), \bm{\Lambda} |  T^A  |g(x), \bm{\Lambda} \right>
  =  \left< \xi(x), \bm{\Lambda} |  T^A  | \xi(x), \bm{\Lambda} \right>
  .
\end{equation}
using   the highest weight state $\left|  \Lambda \right>$ of the respective fundamental representation. 
The state  $\left| \xi(x), \bm{\Lambda} \right>$ is regarded as the coherent state describing the subspace corresponding to the subgroup $G/\tilde{H}=SU(3)/U(2) \simeq CP^2$, the two-dimensional complex projective space.
This is also assured in the following way.
The component $m^A$ of $\bm{m}$ is rewritten as 
\begin{equation}
  m^A(x)  
 = \phi_a^*(x) (T^A)_{ab} \phi_b(x)  \ \ (a,b=1,2,3)
   ,
\end{equation}
by introducing the $CP^2$ variable $\phi^a(x)$:
\begin{equation}
   \phi_a(x)  := (g(x) \left|\bm{\Lambda}  \right>)_a 
   .
\end{equation}
The complex field $\phi^a(x)$ is indeed the $CP^2$ variable, since there are only two independent complex degrees of freedom among three variables $\phi^a(x)$  which are subject to the constraint:
\begin{equation}
  \phi^\dagger (x) \phi(x)
= \sum_{a=1}^{3}  \phi_a^* (x) \phi_a(x)  
=  \left< \bm{\Lambda} \right| g^{\dagger}(x)  g(x) \left|\bm{\Lambda}  \right> 
 =  \left< \bm{\Lambda}   |\bm{\Lambda}  \right>
 = 1
  .
\end{equation}
For more details, see \cite{KT99}.
This result suggests that the Wilson loop operator in fundamental representations of SU(N) can be studied by the $CP^{N-1}$ valued field effectively, rather than $F_{N-1}$.

\section{Non-Abelian Stokes theorem, revisited}\label{sec:NASTr}

Applying (\ref{aaa}) to (\ref{mdef}) and (\ref{omegadef}) in (\ref{NAST0}), we obtain
\footnote{
The factor 2 in front of the trace is due to the normalization
${\rm tr}(T_A T_B)=\frac12 \delta_{AB}$ adopted in this paper.
}  
\begin{align}
 m^A(x)  
=&   2{\rm tr} \left\{ \mathcal{H} \xi^\dagger(x) T^A \xi(x) \right\}
=   2{\rm tr} \left\{ \bm{m}(x) T^A \right\}
 ,
\label{ndef}
\\
  \omega(x)   
=&   -2{\rm tr} \left\{ \mathcal{H} ig^{-1} \xi^\dagger (x) d\xi(x) \right\}
 .
\label{omegadef2}
\end{align}
This leads to another representation of (\ref{NAST0}):
The path ordering $\mathscr{P}$  in the  Wilson loop operator 
\begin{equation}
 W_C[A]  
 := {\rm tr} \left[ \mathscr{P} \exp \left\{ 
ig \oint_{C} \mathscr{A}(x)
\right\} \right] /{\rm tr}(\bm{1}) ,
\quad 
\mathscr{A}(x) := \mathscr{A}_\mu^A(x) T^A  dx^\mu 
  ,
\end{equation}
can be eliminated at the price of introducing integrations over all gauge transformations along the loop $C$:
\begin{equation}
 W_C[A] 
 = \int \mathcal{D}_{C}G  \exp \left\{ 
ig \oint_{C}   2{\rm tr} (\mathcal{H} [ G^{\dagger}(x) \mathscr{A}(x)  G(x) - ig^{-1} G^{\dagger}(x) d G(x) ] )
\right\}  
 ,
\end{equation}
where $\mathcal{D}_{C}G$ is the product of the invariant Haar measure $dG(x)$:
\begin{equation}
 \mathcal{D}_{C}G   := \prod_{x \in C} dG(x)
 .
\end{equation}
Then we can cast the line integral to the surface integral using the usual Stokes theorem:
\begin{equation}
 W_C[A]  = \int \mathcal{D}_{C}G  \exp \left\{ 
ig \oint_{C} A  
\right\}  
= \int \mathcal{D}_{\Sigma}G  \exp \left\{ 
ig  \int_{\Sigma:\partial \Sigma=C} dA  
\right\}  
 , \quad
\end{equation}
where $V$ is the one-form defined by
\begin{equation}
 A := A_\mu dx^\mu 
, \quad 
 A_\mu(x) := 2{\rm tr}(\bm{m}(x) \mathscr{A}_\mu(x))   -  2{\rm tr}(\mathcal{H}ig^{-1}  G^{\dagger}(x) \partial_\mu G(x))  
 .
\end{equation}

Therefore, the Wilson loop operator originally defined in the line integral form along the closed loop $C$ is cast into the surface integral form over the surface $\Sigma$ bounded by $C$. 
This version of the non-Abelian Stokes theorem is called the Diakonov-Petrov (DP) type, since this form was for the first time derived by Diakonov and Petrov for the SU(2) gauge group \cite{DP89}.  
The DP version of the non-Abelian Stokes theorem is regarded as the path integral representation of the Wilson loop operator, which enables us to extend the non-Abelian Stokes theorem to more general gauge groups by using the coherent state representation,  as demonstrated in \cite{KondoIV,KT99}.

The curvature two-form is calculated as
\begin{align}
 F=dA =& \frac12 (\partial_\mu A_\nu - \partial_\nu A_\mu) dx^\mu \wedge dx^\nu 
 = F^{(1)} + F^{(2)}
 .
\end{align}
Here the first term reads
\begin{align}
 (\partial_\mu A_\nu - \partial_\nu A_\mu)^{(1)}(x) 
 = \partial_\mu  2{\rm tr}(\bm{m}(x) \mathscr{A}_\nu(x)) -  \partial_\nu  2{\rm tr}(\bm{m}(x) \mathscr{A}_\mu(x))  
 .
\end{align}

It can be shown \cite{KSM08} that the original SU(N) gauge field $\mathscr{A}_\mu(x)$ is decomposed as
\begin{equation}
 \mathscr{A}_\mu(x)
=\mathscr{V}_\mu(x)+\mathscr{X}_\mu(x)
 ,
\end{equation}
such that  
$\mathscr{V}_\mu(x)$ transforms under the gauge transformation just like the original gauge field $\mathscr{A}_\mu(x)$, while $\mathscr{X}_\mu(x)$ transforms like an adjoint matter field:
\begin{align}
  \mathscr{V}_\mu(x) & \rightarrow \mathscr{V}_\mu^\prime(x) = \Omega(x) (\mathscr{V}_\mu(x) + ig^{-1} \partial_\mu) \Omega^{\dagger}(x) 
 ,
 \label{V-ctransf}
  \\
  \mathscr{X}_\mu(x) & \rightarrow \mathscr{X}_\mu^\prime(x) = \Omega(x)  \mathscr{X}_\mu(x) \Omega^{\dagger}(x) 
 \label{X-ctransf}
 , 
\end{align}
by way of a \textit{single} $\bm{m}$ or $\bm{n}$ field which transforms as 
\begin{equation}
 \bm{m}(x)   \rightarrow \bm{m}^\prime(x) = \Omega(x)  \bm{m}(x) \Omega^{\dagger}(x)
 .
\end{equation}
Moreover, the field $\mathscr{V}_\mu(x)$ can be further decomposed as
\begin{equation}
 \mathscr{V}_\mu(x)
=\mathscr{C}_\mu(x)+\mathscr{B}_\mu(x)
 ,
\end{equation}
such that $\mathscr{C}_\mu(x)$ and $\mathscr{B}_\mu(x)$ are the parallel and perpendicular to $\bm{m}(x)$ in the sense that
\begin{align}
 {\rm tr}( \bm{m}(x) \mathscr{C}_\mu(x) ) = {\rm tr}( \bm{m}(x) \mathscr{A}_\mu(x) )    ,
\quad
  {\rm tr}( \bm{m}(x) \mathscr{B}_\mu(x) ) = 0
   ,
\end{align}
in addition to
\begin{align}
  {\rm tr}( \bm{m}(x) \mathscr{X}_\mu(x) ) = 0
   .
\end{align}
The decomposed fields $\mathscr{C}_\mu(x)$,  $\mathscr{B}_\mu(x)$ and $\mathscr{X}_\mu(x)$ are explicitly written in terms of $\mathscr{A}_\mu(x)$ and $\bm{m}(x)$. 
They are regarded as those obtained by the (non-linear) change of variables from the original gauge field. 
For the SU(2) case, this is well-known as the Cho-Faddeev-Niemi-Shabanov (CFNS) decomposition \cite{Cho80,FN98,Shabanov99}. However, SU(N) case needs further discussions. 
See \cite{KSM08} and Appendix~\ref{section:CFNS} for the details. 
Consequently, we obtain
\begin{align}
 (\partial_\mu A_\nu - \partial_\nu A_\mu)^{(1)}(x) 
=  \partial_\mu  2{\rm tr}(\bm{m}(x) \mathscr{C}_\nu(x)) -  \partial_\nu  2{\rm tr}(\bm{m}(x) \mathscr{C}_\mu(x))  
 .
\end{align}

On the other hand, the second term reads
\begin{align}
 (\partial_\mu A_\nu - \partial_\nu A_\mu )^{(2)}(x)
 =&  2{\rm tr}(\mathcal{H} \{ \partial_\mu[ig^{-1}  G^{\dagger}(x) \partial_\nu G(x)] - \partial_\nu[ig^{-1}  G^{\dagger}(x) \partial_\mu G(x)]\} )  
\nonumber\\
 =&   2{\rm tr}(\mathcal{H} \{  ig^{-1}  \partial_\mu G^{\dagger}(x) \partial_\nu G(x)  -  ig^{-1} \partial_\nu G^{\dagger}(x) \partial_\mu G(x) \} )  
\nonumber\\&
 +    2{\rm tr}(\mathcal{H}  ig^{-1}   G^{\dagger}(x) [\partial_\mu , \partial_\nu] G(x)  )  
\nonumber\\
 =&   2{\rm tr}(\mathcal{H} ig   [ig^{-1}   G^{\dagger}(x) \partial_\mu G(x), ig^{-1}  G^{\dagger}(x) \partial_\nu G(x)] )  
 ,
\end{align}
or
\footnote{
Hereafter, we omit the last term
$
  2{\rm tr}(\mathcal{H}  ig^{-1}   G^{\dagger}(x) [\partial_\mu , \partial_\nu] G(x)  )  
$. 
}
\begin{align}
 (\partial_\mu A_\nu - \partial_\nu A_\mu )^{(2)}(x)
 =&   2{\rm tr}(\mathcal{H} ig   [ig^{-1}   G^{\dagger}(x) \partial_\mu G(x), ig^{-1}  G^{\dagger}(x) \partial_\nu G(x)] )  
\nonumber\\
 =&   2{\rm tr}(\bm{m}(x) ig    [ig^{-1}     \partial_\mu G(x) G^{\dagger}(x), ig^{-1}  \partial_\nu G(x)  G^{\dagger}(x)]   )  
\nonumber\\
 =&   2{\rm tr}(\bm{m}(x) ig    [\mathscr{B}_\mu(x), \mathscr{B}_\nu(x)]   )  
\nonumber\\ 
 =&   2{\rm tr}(\bm{m}(x) \mathscr{F}_{\mu\nu}[\mathscr{B}](x)   )  
 .
\end{align}
where we have used
\begin{align}
 \mathscr{B}_\mu(x) 
= -ig^{-1}     \partial_\mu G(x) G^{\dagger}(x)
=  ig^{-1}     G(x)  \partial_\mu G^{\dagger}(x)
 ,
\end{align} 
and
\begin{align}
 \mathscr{F}_{\mu\nu}[\mathscr{B}](x) 
= ig [ \mathscr{B}_\mu(x) , \mathscr{B}_\nu(x) ]  
 .
\end{align} 
Therefore, we have arrived at the final expression:
\begin{align}
 F_{\mu\nu}(x) :=&  \partial_\mu A_\nu(x) - \partial_\nu A_\mu(x)  
 \nonumber\\
  =&  2{\rm tr}(\bm{m}(x) \mathscr{F}_{\mu\nu}[\mathscr{V}](x)   )  
 \nonumber\\
  =&  2{\rm tr}(  \bm{m}(x) \mathscr{F}_{\mu\nu}[\mathscr{C}](x) )
+   2{\rm tr}(  \bm{m}(x) \mathscr{F}_{\mu\nu}[\mathscr{B}](x) )
  \nonumber\\
  =& \partial_\mu 2{\rm tr}(\bm{m}(x) \mathscr{C}_\nu(x)) - 
  \partial_\nu 2{\rm tr}(\bm{m}(x) \mathscr{C}_\mu(x)) 
  + 2{\rm tr}(4ig^{-1} \bm{m}(x) [\partial_\mu \bm{m}(x), \partial_\nu \bm{m}(x)])
   ,
\end{align}
where we have used the relation shown in (\ref{mF}): 
\begin{align}
   {\rm tr}( \bm{m}(x)\mathscr{F}_{\mu\nu}[\mathscr{B}](x) )
=   4
 {\rm tr}(ig^{-1} \bm{m}(x) [\partial_\mu \bm{m}(x), \partial_\nu \bm{m}(x)])
 .
\end{align}
Thus the Wilson loop operator originally defined in terms of $\mathscr{A}_\mu(x)$ has been rewritten in terms of only the variable $\mathscr{V}_\mu(x)$  and the variable $\mathscr{X}_\mu(x)$ has disappeared in the final expression: 
\begin{equation}
 W_C[A] 
 = \int \mathcal{D}_{\Sigma}G  \exp \left\{ 
ig  \int_{\Sigma:\partial \Sigma=C} 2{\rm tr}(\bm{m}(x)\mathscr{F}[\mathscr{V}](x)   )  
\right\}  
 , \quad
 \bm{m}(x)  
  =  G(x) \mathcal{H} G(x)^\dagger
  .
\end{equation}
 This is the gauge-invariant manifestation of the Abelian dominance for SU(N) Yang-Mills gauge theory, which extends the SU(2) case obtained in \cite{Cho00}.
More details on the Abelian dominance will be discussed in \cite{KS08}.

For the SU(2) Wilson loop operator in the fundamental representation, $\bm{m}$ field agrees with the color field $\bm{n}$ up to a numerical factor:
\begin{equation}
 \bm{m}(x) = \frac{-1}{2} \bm{n}(x) 
  ,
\end{equation}
which reproduces the well-known field strength of the form:
\begin{align}
 F_{\mu\nu}(x) =&  \frac{-1}{2}
  {\rm tr}( 2\bm{n}(x) \mathscr{F}_{\mu\nu}[\mathscr{V}](x) )
  \nonumber\\
  =& \frac{-1}{2} [\partial_\mu 2{\rm tr}(\bm{n}(x) \mathscr{C}_\nu(x)) - 
  \partial_\nu 2{\rm tr}(\bm{n}(x) \mathscr{C}_\mu(x)) 
  + 2{\rm tr}(ig^{-1} \bm{n}(x) [\partial_\mu \bm{n}(x), \partial_\nu \bm{n}(x)]) ]
   .
\end{align}
This is rewritten into another manifestly gauge-invariant form by using the covariant derivative 
$
 D_\mu^{[\mathscr{A}]} \bm{n}(x) := \partial_\mu \bm{n}(x) -ig [ \mathscr{A}_\mu(x), \bm{n}(x) ] 
$:
\begin{align}
 F_{\mu\nu}(x)  
= \frac{-1}{2}
  2{\rm tr}\{ \bm{n}(x) \mathscr{F}_{\mu\nu}[\mathscr{A}](x) 
  +  ig^{-1} \bm{n}(x) [D_\mu^{[\mathscr{A}]} \bm{n}(x), D_\nu^{[\mathscr{A}]} \bm{n}(x)] \}
   ,
\end{align}
since the gauge transformation is given by
\begin{align}
  \bm{n}(x) 
& \rightarrow \Omega(x) \bm{n}(x) \Omega^\dagger(x)
   ,
\nonumber\\
  \mathscr{F}_{\mu\nu}[\mathscr{A}](x) 
& \rightarrow \Omega(x) \mathscr{F}_{\mu\nu}[\mathscr{A}](x) \Omega^\dagger(x)
   ,
\nonumber\\
   D_\mu^{[\mathscr{A}]} \bm{n}(x) 
& \rightarrow \Omega(x) D_\mu^{[\mathscr{A}]} \bm{n}(x) \Omega^\dagger(x)
   .
\end{align}
Under the identification between the color field and the (normalized) Higgs field:
\begin{equation}
  n^A(x) \leftrightarrow 
\hat{\phi}^A(x):=\phi^A(x)/|\bm{\phi}(x)|  
\quad 
(|\bm{\phi}(x)| := \sqrt{2{\rm tr}\{\bm{\phi}(x)\bm{\phi}(x)\}} )
   ,
\end{equation}
the antisymmetric tensor of rank two 
$f_{\mu\nu}=2F_{\mu\nu}$ in the Yang-Mills theory has the same form as the 't Hooft--Polyakov tensor in the Georgi-Glashow model which leads to the SU(2) gauge-invariant 't Hooft--Polyakov magnetic monopole.
This fact suggests that the gauge-invariant magnetic monopole is defined even in the pure Yang-Mills theory without the Higgs field, as shown explicitly in the next section.

For  SU(3) in the fundamental representation, the simplest choice is 
\begin{equation}
 \bm{m}(x) = \frac{-1}{\sqrt{3}} \bm{h}(x) \quad \bm{h}(x) := \bm{n}_2(x) 
= G(x) \frac{\lambda_8}{2}  G^\dagger(x) 
  .
\end{equation}
This leads to the field strength
\begin{align}
 F_{\mu\nu}(x)
 =&  \frac{-1}{\sqrt{3}}
  {\rm tr}( 2\bm{h}(x) \mathscr{F}_{\mu\nu}[\mathscr{V}](x) )
  \nonumber\\
  =& \frac{-1}{\sqrt{3}} [\partial_\mu 2{\rm tr}(\bm{h}(x) \mathscr{C}_\nu(x)) - 
  \partial_\nu 2{\rm tr}(\bm{h}(x) \mathscr{C}_\mu(x)) 
  + 2{\rm tr}(\frac{4}{3}ig^{-1} \bm{h}(x) [\partial_\mu \bm{h}(x), \partial_\nu \bm{h}(x)]) ]
   .
\end{align}

For  SU(N) in the fundamental representation, we can choose  
\begin{equation}
 \bm{m}(x) = - \sqrt{\frac{N-1}{2N}} \bm{h}(x) , \quad \bm{h}(x) 
:= \bm{n}_{N-1}(x) 
= G(x) H_{N-1}  G^\dagger(x) 
  .
\end{equation}
This leads to the field strength
\begin{align}
 F_{\mu\nu}(x) =&  - \sqrt{\frac{N-1}{2N}}
  {\rm tr}( 2\bm{h}(x) \mathscr{F}_{\mu\nu}[\mathscr{V}](x) )
  \nonumber\\
  =& - \sqrt{\frac{N-1}{2N}} \{ \partial_\mu 2{\rm tr}(\bm{h}(x) \mathscr{C}_\nu(x)) - 
  \partial_\nu 2{\rm tr}(\bm{h}(x) \mathscr{C}_\mu(x)) 
  \nonumber\\
  &+ 2{\rm tr}(\frac{2(N-1)}{N}ig^{-1} \bm{h}(x) [\partial_\mu \bm{h}(x), \partial_\nu \bm{h}(x)]) \} 
   .
   \label{F-def}
\end{align}

Finally, we consider the Abelian limit $G \rightarrow U(1)^{N-1}$ and the Abelian case $G=U(1)$.  
In the Abelian case $G=U(1)$, we  need neither taking the trace nor inserting  the complete sets. 
The Haar measure disappears from the representation. 
The off-diagonal elements do not exist.  Therefore, the non-Abelian Stokes theorem reproduces the usual Stokes theorem.

\section{Magnetic monopole and Wilson loop}\label{sec:magnetic-monopole}

\par
Let $\sigma=(\sigma^0,\sigma^1)=(\tau,\sigma)$ be the world sheet coordinates on the two-dimensional surface $\Sigma_C$ which is bounded by the Wilson loop $C$, while let  $x(\sigma)$  be the target space coordinate of the surface $\Sigma_C$ in $\mathbb{R}^D$.  
First of all, we rewrite the surface integral 
$\int_{\Sigma_C} dS^{\mu\nu} F_{\mu\nu}$
 into the volume integral:
\begin{align}
\int_{\Sigma_C:\partial \Sigma_C=C} F
 =& 
 \int_{\Sigma_C} dS^{\mu\nu}(x(\sigma)) F_{\mu\nu}(x(\sigma)) 
\nonumber\\
 =& \int d^Dx F_{\mu\nu}(x) \Theta_{\mu\nu}^{\Sigma_C}(x)  ,
\end{align}
where we have introduced 
an antisymmetric tensor of rank two,
\begin{align}
 \Theta_{\mu\nu}^{\Sigma_C}(x) 
:=   \int_{\Sigma_C}  dS^{\mu\nu}(x(\sigma)) \delta^D(x-x(\sigma)) 
 .
\end{align}
We call $\Theta_{\mu\nu}^{\Sigma_C}(x)$ the vorticity tensor  with the support on the surface $\Sigma_C$ spanned by the Wilson loop $C$.  
Here the surface element $dS^{\mu\nu}$ of $\Sigma_C$ is rewritten using the Jacobian $J$ from $x^\mu, x^\nu$ to $\sigma_1,\sigma_2=\tau,\sigma$ as
\begin{align}
dS^{\mu\nu}(x(\sigma)) =& {1 \over 2}  d^2 \sigma \epsilon^{ab}{\partial x^\mu \over \partial \sigma^a}{\partial x^\nu \over \partial \sigma^b} 
= {1 \over 2} d^2 \sigma J^{\mu\nu}(\sigma) ,
\quad 
 J^{\mu\nu}(\sigma) :=  \epsilon^{ab}{\partial x^\mu \over \partial \sigma^a}{\partial x^\nu \over \partial \sigma^b} 
= {\partial(x^\mu,x^\nu) \over \partial(\sigma^0,\sigma^1)} 
 .
\end{align}

Second, it is rewritten in terms of two conserved currents, the  ``magnetic-monopole current''   $k$ and the ``electric current'' $j$, defined by
\begin{align}
 k:=& \delta *f = *df,
 \nonumber\\
 j:=& \delta f 
  ,
\end{align}
where  $f$ is the two-form defined by $f=\sqrt{\frac{2N}{N-1}} F$ (\ref{F-def}):
\begin{align}
 f_{\mu\nu}(x) 
  =&  \partial_\mu 2{\rm tr}(\bm{h}(x) \mathscr{A}_\nu(x)) - 
  \partial_\nu 2{\rm tr}(\bm{h}(x) \mathscr{A}_\mu(x)) 
  \nonumber\\
  &+ 2{\rm tr}(\frac{2(N-1)}{N}ig^{-1} \bm{h}(x) [\partial_\mu \bm{h}(x), \partial_\nu \bm{h}(x)])  
   .
\end{align}
In fact, we find 
\begin{align}
\int_{\Sigma:\partial \Sigma=C} f
 =&  \int d^Dx \Theta_{\Sigma}^{\mu\nu}(x) f_{\mu\nu}(x)
\nonumber\\
:=& (\Theta_{\Sigma},f) 
\nonumber\\
=& (*\Theta_{\Sigma},*f) 
\nonumber\\
=& (*\Theta_{\Sigma},\Delta^{-1}(d\delta+\delta d)*f)
\nonumber\\
=& (*\Theta_{\Sigma},\Delta^{-1}d\delta *f) + (*\Theta_{\Sigma},\Delta^{-1}\delta d*f)
\nonumber\\
=& (\delta \Delta^{-1} *\Theta_{\Sigma}, \delta *f) + (\Theta_{\Sigma}, *\Delta^{-1}\delta * \delta f) 
\nonumber\\
=& (\delta \Delta^{-1} *\Theta_{\Sigma}, \delta *f) + (\Theta_{\Sigma} ,\Delta^{-1}d  \delta f)  
\nonumber\\
=& (\delta \Delta^{-1} *\Theta_{\Sigma}, k) + (\Delta^{-1}\delta \Theta_{\Sigma},  j)  ,
\end{align}
where we have used $**=(-1)^{p(D-p)}$ and $\delta=(-1)^p *d*$ ($D=$odd), $\delta=- *d*$ ($D=$even) when applied to $p$-form in $D$-dimensional Euclidean space and 
 $\Delta$ is the $D$-dimensional Laplacian (d'Alembertian)$\Delta:=d\delta+\delta d$.

In this way we obtain another expression of the NAST for the Wilson loop operator in the fundamental representation of SU(N):
\begin{equation}
 W_C[A] 
= \int [d\mu(\xi)]_{\Sigma} \exp \left\{ ig  \sqrt{\frac{N-1}{2N}} (k, \Xi_{\Sigma}) + ig  \sqrt{\frac{N-1}{2N}}  (j, N_{\Sigma}) \right\} ,
\label{NAST-SUN}
\end{equation}
where $\Xi_{\Sigma}$ and $N_{\Sigma}$ are defined by
\begin{equation}
 \Xi_{\Sigma} := * d\Theta_{\Sigma} \Delta^{-1} = \delta *\Theta_{\Sigma} \Delta^{-1} , \quad
 N_{\Sigma} := \delta \Theta_{\Sigma} \Delta^{-1} .
\end{equation}
For SU(2), in particular, arbitrary representation is characterized by $J=\frac12, 1, \frac32, 2, \frac52, \cdots$. The Wilson loop operator in the representation $J$ for SU(2) obey the non-Abelian Stokes theorem:
\begin{equation}
 W_C[A] 
= \int [d\mu(\xi)]_{\Sigma} \exp \left\{ ig J  (k, \Xi_{\Sigma}) + ig J  (j, N_{\Sigma}) \right\} 
.
\end{equation}
This agrees with (\ref{NAST-SUN}) for a fundamental representation $J=\frac12$ of SU(2). 
Thus, the Wilson loop can  be expressed by the electric current $j_\mu$ and the magnetic monopole current $k_\mu$ which depend on the group $G$, while 
$N_{\Sigma}$ and $\Xi_{\Sigma}$ do not depend on the group $G$ and depend only on the geometry of the surface $\Sigma$. 

Note that $k$ is $(D-3)$-form and $j$ is one-form in $D=d+1$ dimensions.   
$N_{\Sigma}$ is one-form in any dimension 
having the component:
\begin{align}
 N_{\Sigma}^\mu(x)  =& \partial_\nu^x \int d^4y \Theta_{\Sigma}^{\mu\nu}(y) \Delta_{(D)}^{-1}(y-x)
\nonumber\\
 =& {1 \over 2} \partial_\nu^x \int_{\Sigma} d^2 S^{\mu\nu}(x(\sigma)) \Delta_{(D)}^{-1}(x(\sigma)-x) .
\end{align}
Whereas, $\Xi_{\Sigma}$ is $(D-3)$-form for $D=d+1$ dimensional case. The explicit form is obtained by using the $D$-dimensional Laplacian (d'Alembertian) $\Delta_{(D)}$ in the spacetime dimension $D$ in question as follows.

We show below that the factor 
$W_C^m:=\exp [ig   \sqrt{\frac{N-1}{2N}}  (k, \Xi_{\Sigma})]$ has geometrical and topological meanings.

For $D=3$, $\Xi_{\Sigma}$ is the zero-form with the component:
\begin{align}
 \Xi_{\Sigma}(x)  =& {1 \over 2} \epsilon^{\nu\rho\sigma} \partial_\nu^x \int d^4y \Theta^{\Sigma}_{\rho\sigma}(y) \Delta_{(3)}^{-1}(y-x)
\nonumber\\
=& {1 \over 2} \epsilon^{\nu\rho\sigma}  \int_{\Sigma_C} d^2 S_{\rho\sigma}(x(\sigma)) \partial_\nu^x \Delta_{(3)}^{-1}(x(\sigma)-x)  
\nonumber\\
=& {1 \over 2} \epsilon^{\nu\rho\sigma} \int_{\Sigma_C} d^2 S_{\rho\sigma}(x(\sigma)) \partial_\nu^x  \frac{1}{4\pi|x-x(\sigma)|} ,
\end{align}
while
 $k$ is also zero-form, i.e, the magnetic monopole density function:
\begin{equation}
 k = \rho_m = {1 \over 2} \epsilon^{\nu\rho\sigma}  \partial_\nu f_{\rho\sigma} .
\end{equation}

\begin{figure}[ptb]
\begin{center}
\includegraphics[height=1.5in]{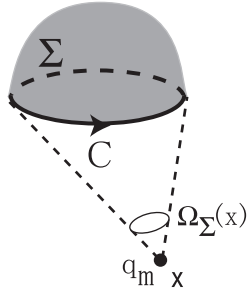}
\end{center} 
 \caption[]{
 The magnetic monopole $q_m$ at $x$ and the solid angle $\Omega_{\Sigma}(x)$ at $x$ subtended by the surface $\Sigma$ bounding the Wilson loop $C$.
}
 \label{fig:Wilson-solid-angle}
\end{figure}

It is known that the solid angle $\Omega_{\Sigma}(x)$ under which the surface $\Sigma$ shows up to an observer at the point $x$ is written as
\begin{align}
 \Omega_{\Sigma}(x) 
=& \frac{\partial}{\partial x^\mu} \int_{\Sigma_C} d^2S^\mu(y) \frac{1}{|x-y|}   
= \frac{1}{2} \epsilon^{\mu\alpha\beta}  \frac{\partial}{\partial x^\mu} \int_{\Sigma_C} d^2S_{\alpha\beta}(y) \frac{1}{|x-y|} 
\nonumber\\
 =& 4\pi \Xi_{\Sigma}(x) , 
\end{align}
where
\begin{align}
 d^2S^\mu := \frac{1}{2} \epsilon_{\mu\alpha\beta} d^2S_{\alpha\beta}  .
\end{align}
In the case when $\Sigma$ is a closed surface surrounding the point $x$, we have $\Omega_{\Sigma}=4\pi$, since due to the Gauss law
\begin{align}
 \Omega_{\Sigma}(x) 
=& \oint_{\Sigma} d^2S^\mu(y)  \frac{\partial}{\partial x^\mu}  \frac{1}{|x-y|} 
\nonumber\\
=& \int_{V:\partial V=\Sigma} d^3y  \frac{\partial}{\partial x_\mu}  \frac{\partial}{\partial x^\mu}  \frac{1}{|x-y|} 
\nonumber\\
=& \int_{V} d^3y  4\pi \delta^3(x-y)  
= 4\pi
  .  
\end{align}
This is a standard result for the total solid angle in three dimensions.

Thus, for $D=3$, $\Xi_{\Sigma}$ is the normalized solid angle  ($\Omega_{\Sigma}$ divided by the total solid angle $4\pi$) and the exponential factor in SU(2) NAST reads 
\begin{equation}
W_C^m = \exp \left[ i g \frac12 \int d^3x k(x) \Xi_{\Sigma}(x) \right]
= \exp \left[ i g \frac12 \int d^3x \rho_m(x) \frac{\Omega_{\Sigma}(x)}{4\pi} \right] .
\end{equation}
See Fig.~\ref{fig:Wilson-solid-angle}.

\begin{figure}[ptb]
\begin{center}
\includegraphics[height=1in]{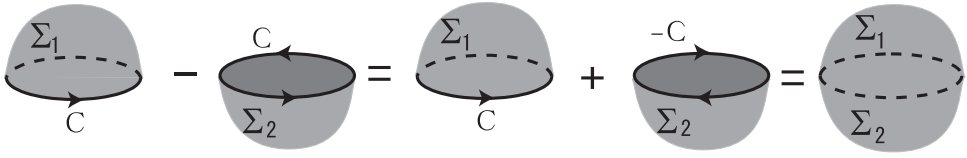}
\end{center} 
 \caption[]{
 Quantization of the magnetic charge.  The difference between  the surface integrals of $f$ over two surfaces $\Sigma_{1}$ and $\Sigma_{2}$ with the same boundary $C$ is equal to the surface integral $\int_{\Sigma} f$ of $f$ over one closed surface $\Sigma=\Sigma_{1}+\Sigma_{2}$. 
Here the direction of the normal vector to the surface $\Sigma_{2}$ must be consistent with the direction of the line integral over $C$. 
 The magnetic charge $Q_m$ is non-zero if the magnetic monopole exists inside $\Sigma$, otherwise it is zero. 
}
 \label{fig:Wilson-solid-angle2}
\end{figure}

We examine the relationship between the magnetic charge and its quantization condition. In order to extract the information on the magnetic charge through the non-Abelian Stokes theorem for the Wilson loop operator, we must consider the integration of the curvature two-form $f$ over the closed surface $\Sigma$.   
In fact, the Dirac quantization condition $q_m=4\pi g^{-1} n$ for the magnetic charge $q_m$ is obtained for SU(2) from the condition of the non-Abelian Stokes them does not depend on the surface chosen for spanning the surface bounded by the loop $C$, remembering that the original Wilson loop is defined for the specified closed loop $C$.
\begin{align}
2\pi n 
=& 
g \frac12 \int d^3x \rho_m(x) \frac{\Omega_{\Sigma_{1}}(x)}{4\pi} 
-  g \frac12 \int d^3x \rho_m(x) \frac{\Omega_{\Sigma_{2}}(x)}{4\pi} 
\nonumber\\
=& g \frac12 \int d^3x \rho_m(x) \frac{\Omega_{\Sigma_{1}}(x)-\Omega_{\Sigma_{2}}(x)}{4\pi} 
\nonumber\\
=& g \frac12 \int d^3x \rho_m(x)  
\nonumber\\
=& g \frac12 q_m 
 ,
\end{align}
where we have used
$\Omega_{\Sigma_{1}}(x)-\Omega_{\Sigma_{2}}(x)=4\pi$.
See Fig.~\ref{fig:Wilson-solid-angle2}.

For an ensemble of point-like magnetic charges located at $x=z_a$ ($a=1, \cdots, n$)
\begin{equation}
 k(x) = \rho_m(x) = \sum_{a=1}^{n} q_m^a \delta^{(3)}(x-z_a) 
  ,
  \quad 
  q_m^a=4\pi g^{-1} n_a , \quad n_a \in \mathbb{Z} 
   ,
\end{equation}
we have a geometric representation:
\begin{equation}
W_C^m  = \exp \left\{ i \frac12 \frac{g}{4\pi} \sum_{a=1}^{n} q_m^a \Omega_{\Sigma}(z_a)  \right\}
= \exp \left\{ i \frac12  \sum_{a=1}^{n} n_a \Omega_{\Sigma}(z_a)  \right\}
 , \quad n_a \in \mathbb{Z} 
  .
\end{equation}

For $D=4$, the magnetic current one-form $k$ in the continuum   SU(N) Yang-Mills theory is defined as
\begin{equation}
 k^\mu = \frac12 \epsilon^{\mu\nu\rho\sigma} \partial_\nu f_{\rho\sigma} 
   .
\end{equation}
The magnetic current $k$ is conserved, $\partial_\mu k^\mu=0$. 
Then the magnetic charge is defined by  
\begin{equation}
  Q_m = \int d^3x k^0 
= \int d^3x \frac12 \epsilon^{jk\ell} \partial_\ell    f_{jk}(x) 
   .
   \label{m-charge}
\end{equation}
Whereas, for $D=4$, $\Xi_{\Sigma}$ is one-form with the component:
\begin{align}
 \Xi^\mu_{\Sigma}(x)  =& {1 \over 2} \epsilon^{\mu\nu\rho\sigma} \partial_\nu^x \int d^4y \Theta_{\rho\sigma}(y) \Delta_{(4)}^{-1}(y-x)
\nonumber\\
=& {1 \over 2} \epsilon^{\mu\nu\rho\sigma}  \int_S d^2 S_{\rho\sigma}(x(\sigma)) \partial_\nu^x \Delta_{(4)}^{-1}(x(\sigma)-x) 
\nonumber\\
=&  {1 \over 2} \epsilon^{\mu\nu\rho\sigma}  \int_S d^2 S_{\rho\sigma}(x(\sigma)) \partial_\nu^x \frac{1}{4\pi^2|x-x(\sigma)|^2} 
\nonumber\\
=&    \epsilon^{\mu\nu\rho\sigma}  \int_S d^2 S_{\rho\sigma}(x(\sigma))  \frac{(x(\sigma)-x)_\nu}{4\pi^2|x(\sigma)-x|^4} 
 .
\end{align}

\begin{figure}[ptb]
\begin{center}
\includegraphics[height=1in]{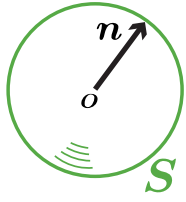}
\end{center} 
 \caption[]{
The color field $\bm{n}$ in SU(2) case.
}
 \label{fig:color-field}
\end{figure}

For SU(2), the rank is one $r=1$. 
\begin{equation}
  \Lambda = J, \quad H = \frac12 \sigma_3 , \quad
  \mathcal{H} = J \frac{\sigma_3}{2} 
  =   J {\rm diag}(\frac12, -\frac12) , \quad
  \bm{m} = J G^{\dagger} \frac{\sigma_3}{2} G 
   .
\end{equation}
Thus we recover the SU(2) case investigated so far \cite{DP89,KondoIV}. 
The SU(2) gauge-invariant magnetic current is obtained from the SU(2) gauge-invariant field strength:
\begin{align}
 f_{\mu\nu} 
 =& 2{\rm tr}( \bm{n}  \mathscr{F}_{\mu\nu}[\mathscr{V}] )
\nonumber\\
=& \partial_\mu c_\nu  - \partial_\nu c_\mu   -  g^{-1}  \mathbf{n}  \cdot  (\partial_\mu \mathbf{n} \times \partial_\nu \mathbf{n} ) 
=    E_{\mu\nu}+H_{\mu\nu} 
 .
\label{latCFN-monop}
\end{align}
We can show that the color field generates the magnetic charge subject to the Dirac quantization condition. 
The color field $\mathbf{n}$ parameterized by two polar angles $(\alpha,\beta)$ on the target space $S^2 \simeq SU(2)/U(1)$, see Fig.~\ref{fig:color-field}:
\begin{align}
   \mathbf{n}(x) 
 :=& \begin{pmatrix} n_1(x) \cr n_2(x) \cr n_3(x) \end{pmatrix}
:= \begin{pmatrix} \sin \beta(x) \cos \alpha(x) \cr
 \sin \beta(x) \sin \alpha(x) \cr
\cos \beta(x)  
 \end{pmatrix}  ,
\end{align}
yields 
\begin{align}
\mathbf{n} \cdot (\partial_\mu \mathbf{n}   \times \partial_\nu \mathbf{n}) 
 = \sin \beta (\partial_\mu \beta \partial_\nu \alpha
 - \partial_\mu \alpha \partial_\nu \beta) 
 = \sin \beta {\partial(\beta, \alpha) 
 \over \partial(x^\mu,x^\nu)}  .
\end{align}
Taking into account the fact that 
${\partial(\beta, \alpha) \over \partial(x^\mu,x^\nu)}$ 
is the Jacobian from $(x^\mu,x^\nu) \in S^2_{phy}$ to $(\beta, \alpha) \in S^2_{int} \simeq SU(2)/U(1)$ parameterized
by $(\beta, \alpha)$, we obtain the Dirac quantization condition:
\begin{align}
 Q_m 
=& \int d^3x \frac12 \partial_\ell \epsilon^{\ell jk}     f_{jk} 
 \nonumber\\
=& \oint_{S^2_{phy}} dS_{\ell} \frac12 \epsilon^{\ell jk}  f_{jk} 
 \nonumber\\
=&  \oint_{S^2_{phy}} dS_{jk} 
  g^{-1} \mathbf{n} \cdot (\partial_j \mathbf{n}   \times \partial_k \mathbf{n}) 
 \nonumber\\
=& -g^{-1} \oint_{S^2_{phy}} dS_{jk} 
  {\partial(\beta, \alpha)  \over \partial(x^j,x^k)} \sin \beta
 \nonumber\\
 =& - g^{-1}  \oint_{S^2_{int}}  d\beta  d\alpha \sin \beta
 \nonumber\\
 =&  4\pi g^{-1}   n  \quad (n=0, \pm1, \cdots)
 ,
\end{align}
since $d\beta  d\alpha \sin \beta$ is the surface element on $S^2_{int}$ and a surface of a unit radius has the area $4\pi$. 
Hence $n$ gives a number of times  $S^2_{int}$ is
wrapped by a mapping from  $S^2_{phys} $ to 
$S^2_{int}$. 
This fact is understood as the Homotopy group:
$\Pi_2(SU(2)/U(1)) = \Pi_2(S^2) = \mathbb{Z}$.

For SU(2), the explicit configuration yielding the non-zero magnetic monopole is given as follows. 
We consider the Wu-Yang configuration 
\begin{equation}
n_A(x)=x_A/r , \quad r:=|x|=\sqrt{x_A x_A} \ ( A=1,2,3 )
   .
\end{equation}
This leads to the magnetic-monopole density 
\begin{align}
 \rho_m =& \frac12 \epsilon^{jk\ell} \partial_\ell  H_{jk}
\nonumber\\
=& \frac12 \epsilon^{jk\ell} \partial_\ell 
\left[ -g^{-1} \epsilon^{jkm} \frac{x^m}{r^3} \right] 
\nonumber\\
=&     -g^{-1}  \partial_\ell 
\left[   \frac{x^\ell}{r^3} \right] 
\nonumber\\
=&     - 4\pi g^{-1} \delta^3(x) .
\end{align}
This corresponds to a magnetic monopole with a unit magnetic charge $q_m=4\pi g^{-1}$ located at the origin. 
Hence, we obtain
\begin{align}
W_C^m
=& \exp \left[ -i g \frac12 \int d^3x  4\pi g^{-1} \delta^3(x) \frac{\Omega_{\Sigma}(x)}{4\pi} \right]  
\nonumber\\
=& \exp \left[ -i \frac12  \Omega_{\Sigma}(0)  \right] .
\end{align}
Therefore,  $\exp [ig \frac12  (k, \Xi_{\Sigma})]$ gives a non-trivial factor $\exp [ \pm i \pi]=-1$ for a magnetic monopole with a unit magnetic charge $q_m=4\pi g^{-1}$ at the origin, since 
$\Omega_{\Sigma}(0)=\pm 2\pi$ for the upper or lower hemisphere $\Sigma$. 
Indeed, this result does not depend on which surface bounding $C$ is chosen in the non-Abelian Stokes theorem.

For $D=4$,   $\Xi$ agrees with the four-dimensional solid angle  given by
\begin{equation}
 \Omega^\mu_{\Sigma}(x) 
= \frac{1}{8\pi^2} \epsilon^{\mu\nu\rho\sigma} \frac{\partial}{\partial x^\nu} \int_{\Sigma} d^2S_{\rho\sigma}(y)    \frac{1}{|x-y|^2} 
= \Xi_{\Sigma}^\mu(x) .
\end{equation}
Consequently, for $D=4$ we have 
\begin{equation}
 W_C^m =  \exp \left[ i g \frac12 \int d^4x k_\mu(x) \Xi^\mu_{\Sigma}(x) \right]
=  \exp \left[ i g \frac12 \int d^4x k_\mu(x) \Omega_{\Sigma}^\mu(x) \right].
\end{equation}

For an ensemble of magnetic monopole loops $C^\prime_a$ ($a=1,\cdots,n$):
\begin{equation}
 k^\mu(x) = \sum_{a=1}^{n} q_m^a \oint_{C^\prime_a} dy^\mu_a \delta^{(4)}(x-x_a) , 
 \quad 
 q_m^a = 4\pi g^{-1} n_a 
  ,
\end{equation}
we obtain
\begin{equation}
 W_C^m    
= \exp \left\{ i \frac12 g \sum_{a=1}^{n} q_m^a L(C^\prime_a, \Sigma)  \right\}
= \exp \left\{  2\pi i \sum_{a=1}^{n} n^a L(C^\prime_a, \Sigma)  \right\}
 , 
\quad n_a \in \mathbb{Z} 
  ,
\end{equation}
where
$L(C^\prime,\Sigma)$ is the linking number between the curve $C^\prime$ and the surface $\Sigma$ \cite{HG90}:
\begin{equation}
  L(C^\prime, \Sigma)  
  := \oint_{C^\prime} dy^\mu(\tau) \Xi^\mu_{\Sigma}(y(\tau)) 
  ,
\end{equation} 
where the curve $C$ is identified with the trajectory of a magnetic monopole and the surface $\Sigma$ with the world sheet of a hadron string for a quark-antiquark pair.

\section{SU(3) magnetic charge and quantization condition}\label{sec:magnetic-charge}

By remembering the relationship:
\begin{align}
 F_{\mu\nu} 
=    {\rm tr}( 2\bm{m}(x) \mathscr{F}_{\mu\nu}[\mathscr{V}](x) )
= - \sqrt{\frac{N-1}{2N}}
  {\rm tr}( 2\bm{h}(x) \mathscr{F}_{\mu\nu}[\mathscr{V}](x) )
= - \sqrt{\frac{N-1}{2N}} f_{\mu\nu}(x)
 ,
\end{align}
the magnetic charge is given by
\begin{align}
  Q_m 
=& \int d^3x k^0 
\nonumber\\
=& \int d^3x \frac12 \epsilon^{jk\ell} \partial_\ell    f_{jk}(x) 
\nonumber\\
=&  \int d^3x \frac12 \epsilon^{jk\ell} \partial_\ell (\bm{h}(x) , \mathscr{F}_{jk}[\mathscr{V}](x))
\nonumber\\
=&  \sqrt{\frac{2N}{N-1}} \int d^3x \frac12 \epsilon^{jk\ell} \partial_\ell ({\bm m}(x) , \mathscr{F}_{jk}[\mathscr{V}](x))
\nonumber\\
=&   \int d^3x \frac12 \epsilon^{jk\ell} \partial_\ell ({\bm n}(x) , \mathscr{F}_{jk}[\mathscr{V}](x))
   .
   \label{charge-fs}
\end{align}
The magnetic charge $Q_m$ appears in the factor $W_C^m $ in the Wilson loop operator for the closed surface $\Sigma$ for which $\Omega_{\Sigma}(x)=4\pi$ as 
\begin{equation}
W_C^m  
\rightarrow \exp \left[ i g \sqrt{\frac{N-1}{2N}}  Q_m  \right]= 1
 .
\end{equation}

In the case of SU(3), it is known that one can define two gauge-invariant conserved charges $Q^{(1)}$ and $Q^{(2)}$ from the respective color field $\bm{n}_1$ and $\bm{n}_2$,  which obey the different quantization conditions \cite{EW76,GNO77,Sinha76,KKP82,Shnir05,WY07}:
\begin{align} 
 Q^{(1)} 
:=& \int d^3x \frac12 \epsilon^{jk\ell} \partial_\ell ({\bm n}_1(x) , \mathscr{F}_{jk}[\mathscr{V}](x))
= \frac{2\pi}{g} (2n-n') , 
\nonumber\\
 Q^{(2)} 
:=& \int d^3x \frac12 \epsilon^{jk\ell} \partial_\ell ({\bm n}_2(x) , \mathscr{F}_{jk}[\mathscr{V}](x))
= \frac{2\pi}{g} \sqrt{3} n' , \quad n, n' \in \mathbb{Z} 
  ,
\label{qc}
\end{align}
where 
$\sqrt{3} \bm{m}(x)=\bm{n}(x)=(\cos \vartheta(x)) \bm{n}_1(x)+(\sin \vartheta(x)) \bm{n}_2(x)$.
However, we have shown that the Wilson loop in the fundamental representation does not distinguish $Q^{(1)}$ and $Q^{(2)}$, and can probe only the specific combinations  $Q_m$  represented through $\bm{m}$ in (\ref{charge-fs}).
For fundamental representations of SU(3), therefore, the magnetic charge $Q_m$ obeys the following quantization condition.  
For $[1,0]$ and $[-1,0]$,  
\begin{equation}
  Q_m 
  = \pm    \left[ \frac{\sqrt{3}}{2} Q^{(1)} +   \frac{1}{2}  Q^{(2)} \right] 
  = \pm  \frac{2\pi}{g} \left[  (2n-n') +     n' \right] \frac{\sqrt{3}}{2} 
  = \pm  \frac{2\pi}{g}  \sqrt{3}  n
   ,
\end{equation}
for $[0,-1]$ and $[0,1]$, 
\begin{equation}
  Q_m 
  = \pm   \left[ -\frac{\sqrt{3}}{2} Q^{(1)} +   \frac{1}{2}  Q^{(2)} \right] 
  = \pm  \frac{2\pi}{g} \left[  -(2n-n') +     n' \right] \frac{\sqrt{3}}{2}
  = \pm  \frac{2\pi}{g}  \sqrt{3}  (n'-n) 
   ,
\end{equation}
and, in particular, for $[-1,1]$ and $[1,-1]$,  
\begin{equation}
  Q_m 
  = \pm   \left[  -  Q^{(2)} \right] 
  = \pm  \frac{2\pi}{g} \left[  -2  n' \right] \frac{\sqrt{3}}{2}
  =  \mp  \frac{2\pi}{g}  \sqrt{3}  n'  
   .
\end{equation}
These quantization conditions for $Q_m$ are reasonable because they guarantee that the Wilson loop operator defined originally by the closed loop $C$ does not depend on the choice of the surface $\Sigma$ bounded by the loop $C$ when rewritten into the surface integral form in the non-Abelian Stokes theorem, just as in the SU(2) case:  
\begin{equation}
 W_C^m = \exp \left\{ ig \frac{1}{\sqrt{3}} Q_m \right\} = 1
  \rightarrow 
 Q_m = 2\pi \sqrt{3} g^{-1} n , \ n \in \mathbb{Z} 
   .
\end{equation}

Thus, we have shown that the SU(N) Wilson loop operator can probe a single  (gauge-invariant) magnetic monopole in  the pure Yang-Mills theory, which can be defined in a gauge invariant way  even in the absence of any scalar field.
The magnetic charge is subject to the quantization condition which is analogous to the Dirac type: 
\begin{equation}
  Q_m=2\pi g^{-1} \sqrt{\frac{2N}{N-1}} n , \quad n \in \mathbb{Z} 
  .
\end{equation}
Therefore, one need not to introduce $N-1$ kinds of magnetic monopoles, which are usually supposed to be deduced from the maximal torus group $U(1)^{N-1}$ of SU(N).
Thus, calculating the Wilson loop average reduces to the summation over the contributions coming from the distribution of magnetic monopole charges or currents with the geometric factor related to the solid angle or the linking number. 
This issue will be discussed in a future publication.

\section*{Acknowledgments}
The author is appreciative of continuous  discussions from Takeharu Murakami, Toru Shinohara and Akihiro Shibata.
Part of this work was done while the author stayed as an invited participants of the programme ``Strong Fields, Integrability and Strings'' held at Isaac Newton Institute for Mathematical Sciences in Cambridge U.K. for the period from 6 Aug to 24 Aug, 2007.
The author would like to thank Nick Dorey and Simon Hands who enabled him to attend the programme for hospitality.
This work is financially supported by Grant-in-Aid for Scientific Research (C) 18540251 from Japan Society for the Promotion of Science
(JSPS).

\appendix
\section{Formulae for Cartan subalgebras}\label{section:Cartan}

The Cartan algebras are written as 
\begin{align}
 H_k =& \frac{1}{\sqrt{2k(k+1)}} {\rm diag}(1, 1, \cdots, 1, -k, 0, \cdots, 0) 
 \nonumber\\
 =&  \frac{1}{\sqrt{2k(k+1)}} \left( \sum_{j=1}^{k} \bm{e}_{jj} - k \bm{e}_{k+1,k+1} \right) 
 ,
\end{align}
where we have defined the matrix $\bm{e}_{AB}$ whose $AB$ element has the value 1 and other elements are zero, i.e.,
$(\bm{e}_{AB})_{ab}=\delta_{Aa} \delta_{Bb}$.  
The the $ab$ element reads
\begin{equation}
 (H_k)_{ab} =  \frac{1}{\sqrt{2k(k+1)}} \left( \sum_{j=1}^{k} \delta_{aj}\delta_{bj} - k \delta_{a,k+1}\delta_{b,k+1}  \right) 
 .
\end{equation}
The unit matrix ${\bf 1}$ with the element $\delta_{ab}$ is written as
\begin{equation}
 {\bf 1} =  {\rm diag}(1, 1, \cdots, 1) 
= \sum_{j=1}^{N} \bm{e}_{jj}
 .
\end{equation}
The product of diagonal generators is decomposed as 
\begin{align}
  H_j H_k = 
  \begin{cases}
  \frac{1}{\sqrt{2j(j+1)}} H_k & (j>k) \cr
  \frac{1}{\sqrt{2k(k+1)}} H_j & (j<k) \cr
  \frac{1}{2N}{\bf 1} + \frac{1-k}{\sqrt{2k(k+1)}} H_k + \sum_{m=k+1}^{N-1}  \frac{1}{\sqrt{2m(m+1)}} H_m & (j=k) \cr
  \end{cases} 
 .
\end{align}
The first and second relations are easily derived from the definition. 
The third relation is derived as follows.
\begin{equation}
 H_k H_k = c_0 {\bf 1} + \sum_{m=1}^{N-1} c_m H_m
 , 
\end{equation}
where
\begin{equation}
 c_m = 2{\rm tr}(H_k H_k H_m) ,  
 \quad 
 c_0 = {\rm tr}(H_k H_k)/{\rm tr}({\bf 1}) 
  ,
\end{equation}
Here the coefficients are calculated as
\begin{align}
  c_m =& 2{\rm tr}\left[  \frac{1}{2k(k+1)} {\rm diag}(1, 1, \cdots, 1,k^2, 0, \cdots, 0)  H_m \right] 
  \nonumber\\
  =& 
  \begin{cases}
  0 & (1 \le m \le k-1) \cr
  \frac{1-k}{\sqrt{2k(k+1)}}  & (m=k) \cr
  \frac{1}{\sqrt{2m(m+1)}}   & (k+1 \le m \le N-1) \cr
  \end{cases} 
 ,
\end{align}
and
\begin{equation}
 c_0 = {\rm tr}(H_k H_k)/{\rm tr}({\bf 1}) 
 = \frac12 \frac1N
 . 
\end{equation}

\section{Decomposition formulae}\label{section:formulae1}

For any Lie algebra valued function $\mathscr{V}(x)$, the identity holds \cite{FN99a,Shabanov02}:
\footnote{
The SU(2) version of this identity is 
\begin{align}
  \bm{v} = \bm{n} (\bm{n} \cdot \bm{v}) - \bm{n} \times (\bm{n} \times \bm{v})
=  \bm{v}_\parallel +  \bm{v}_\perp  ,
\end{align}
which follows from a simple identity,
$
 \bm{n} \times (\bm{n} \times \bm{v}) = \bm{n} (\bm{n} \cdot \bm{v}) - (\bm{n} \cdot \bm{n}) \bm{v}.
$
}
\begin{align}
  \mathscr{V} = \sum_{j=1}^{N-1} \bm{n}_j(\bm{n}_j, \mathscr{V}) +  \sum_{j=1}^{N-1} [\bm{n}_j, [\bm{n}_j, v]]
  = \sum_{j=1}^{N-1} 2{\rm tr}(\mathscr{V} \bm{n}_j)\bm{n}_j +  \sum_{j=1}^{N-1} [\bm{n}_j, [\bm{n}_j, \mathscr{V}]]
 .
\label{idv}
\end{align}
This identity is equivalent to the identity \cite{FN99a}
\begin{align}
 \delta^{AB}  = n_j^A n_j^B  - f^{ACD} n_j^C f^{DEB} n_j^E  .
\end{align}

The identity is proved as follows.
By using the adjoint rotation, $\mathscr{V}'=U\mathscr{V}U^\dagger$, we have only to prove
\begin{align}
  \mathscr{V}' = \sum_{j=1}^{N-1} H_j (H_j, \mathscr{V}') + \sum_{j=1}^{N-1} [H_j, [H_j, \mathscr{V}']] .
\label{idd}
\end{align}
The Cartan decomposition for $\mathscr{V}=\mathscr{V}'$ reads
\begin{align}
  \mathscr{V} 
=  \sum_{k=1}^{N-1} V^k H_k + \sum_{\alpha=1}^{(N^2-N)/2} (W^*{}^{\alpha} \tilde{E}_{\alpha} + W^{\alpha} \tilde{E}_{-\alpha})  ,
\end{align}
where the Cartan basis is given by
\begin{align}
 \vec{H} =& (H_1, H_2, H_3, \cdots, H_{N-1}) = (T^3, T^8, T^{15}, \cdots, T^{N^2-1}) ,
\nonumber\\
  \tilde{E}_{\pm 1} =& {1 \over \sqrt{2}}(T^1 \pm  i T^2) , \quad
  \tilde{E}_{\pm 2} = {1 \over \sqrt{2}}(T^4 \pm  i T^5) ,  
  \nonumber\\ &
  \cdots,  \quad
  \tilde{E}_{\pm (N^2-N)/2} =  {1 \over \sqrt{2}}(T^{N^2-3} \pm i T^{N^2-2}) ,
\end{align}
and the complex field is defined by
\begin{align}
  W^1 =& {1 \over \sqrt{2}}(V^1+ i V^2) , \quad
  W^2 = {1 \over \sqrt{2}}(V^4+ i V^5) , \quad 
  \nonumber\\ &
  \cdots,  \quad
  W^{(N^2-N)/2} = {1 \over \sqrt{2}}(V^{N^2-3}+ i V^{N^2-2})  .
\end{align}

Now we calculate the double commutator as
\begin{align}
&  [H_j, [H_j, \mathscr{V}]]
\nonumber\\
 =& \sum_{j=1}^{N-1} V_k [H_j, [H_j, H_k]] + \sum_{\alpha=1}^{(N^2-N)/2} (W^*{}^{\alpha} [H_j, [H_j,  \tilde{E}_{\alpha}]] + W^{\alpha} [H_j, [H_j,  \tilde{E}_{-\alpha}]]) 
 \nonumber\\
 =& 
  \sum_{\alpha=1}^{(N^2-N)/2} (W^*{}^{\alpha} [H_j, \alpha_j \tilde{E}_{\alpha}] + W^{\alpha} [H_j, -\alpha_j   \tilde{E}_{-\alpha}]) 
 \nonumber\\
 =& 
   \alpha_j  \alpha_j \sum_{\alpha=1}^{(N^2-N)/2} (W^*{}^{\alpha}  \tilde{E}_{\alpha}  + W^{\alpha}  \tilde{E}_{-\alpha} )
 .
\end{align}
On the other hand, we have 
\begin{align}
& (H_j, \mathscr{V}) 
 \nonumber\\
 =&  \sum_{k=1}^{N-1} V_k (H_j, H_k) + \sum_{\alpha=1}^{(N^2-N)/2} (W^*{}^{\alpha} (H_j, \tilde{E}_{\alpha}) + W^{\alpha} (H_j, \tilde{E}_{-\alpha}))  
 \nonumber\\
 =&   V^j  ,
\end{align}
since 
$(H_j, \tilde{E}_{\alpha})={\rm tr}(H_j \tilde{E}_{\alpha})=0$. 
Thus the RHS of (\ref{idd}) reduces to 
\begin{align}
 \sum_{j=1}^{N-1} V_j H_j  + \sum_{j=1}^{N-1} \alpha_j  \alpha_j \sum_{\alpha=1}^{(N^2-N)/2} (W^*{}^{\alpha}  \tilde{E}_{\alpha}  + W^{\alpha}  \tilde{E}_{-\alpha} ) .
\end{align}
This is equal to the Cartan decomposition of $\mathscr{V}$ itself, since 
\begin{equation}
\sum_{j=1}^{N-1} \alpha_j  \alpha_j=1
 . 
\end{equation}

\section{More decomposition formulae}\label{section:formulae2}

For any Lie algebra valued function $\mathscr{M}(x)$, the identity holds:
\begin{align}
  \mathscr{V} 
= \sum_{A=1}^{N^2-1} \mathscr{V}^A T_A
=  \tilde{\mathscr{V}} + \bm{h} (\bm{h},\mathscr{V}) 
+  2\frac{N-1}{N}  [\bm{h} , [\bm{h} , \mathscr{V}]]
 ,
\label{idv2}
\end{align}
where we have defined the matrix $\tilde{\mathscr{V}}$  in which all the elements in the last column and the last raw are zero:
\begin{equation}
 \tilde{\mathscr{V}} = \sum_{A=1}^{(N-1)^2-1} \mathscr{V}^A T_A
= \sum_{A=1}^{(N-1)^2-1} (\mathscr{V},T_A) T_A 
= \sum_{A=1}^{(N-1)^2-1} 2{\rm tr}(\mathscr{V}T_A)T_A
 , 
\end{equation}
or
\begin{align}
  \mathscr{V}  =  \tilde{\mathscr{V}} + 2{\rm tr}(\mathscr{V} \bm{h})\bm{h} 
+  2\frac{N-1}{N}  [\bm{h} , [\bm{h} , \mathscr{V}]]
 .
\label{idv3}
\end{align}
Note that $[\tilde{\mathscr{V}},\bm{h}]=0$.

The Cartan decomposition for $\mathscr{V}_N=v' \in su(N)$ reads
\begin{align}
  \mathscr{V}_N 
=&   \sum_{k=1}^{N-1} V_k H_k + \sum_{\alpha=1}^{(N^2-N)/2} (W^*{}^{\alpha} \tilde{E}_{\alpha} + W^{\alpha} \tilde{E}_{-\alpha})  
\nonumber\\
=& \tilde{\mathscr{V}}_{N} + M_{N-1} H_{N-1} 
+ \sum_{\alpha=[(N-1)^2-(N-1)]/2+1}^{(N^2-N)/2} (W^*{}^{\alpha} \tilde{E}_{\alpha} + W^{\alpha} \tilde{E}_{-\alpha})  
 .
\end{align}

Now we calculate the double commutator ($r=N-1$) as
\begin{align}
&  [H_r, [H_r, \mathscr{V}_N]]
\nonumber\\
 =& \sum_{j=1}^{N-1} V_k [H_r, [H_r, H_k]] + \sum_{\alpha=1}^{(N^2-N)/2} (W^*{}^{\alpha} [H_r, [H_r,  \tilde{E}_{\alpha}]] + W^{\alpha} [H_r, [H_r,  \tilde{E}_{-\alpha}]]) 
 \nonumber\\
 =& 
  \sum_{\alpha=[(N-1)^2-(N-1)]/2+1}^{(N^2-N)/2} (W^*{}^{\alpha} [H_r, \alpha_r \tilde{E}_{\alpha}] + W^{\alpha} [H_r, -\alpha_r   \tilde{E}_{-\alpha}]) 
 \nonumber\\
 =& 
   \alpha_r  \alpha_r \sum_{\alpha=[(N-1)^2-(N-1)]/2+1}^{(N^2-N)/2} (W^*{}^{\alpha}  \tilde{E}_{\alpha}  + W^{\alpha}  \tilde{E}_{-\alpha} )
 .
\end{align}
On the other hand, we have 
\begin{align}
& (H_r, \mathscr{V}_N) 
 \nonumber\\
 =&  \sum_{k=1}^{N-1} V_k (H_r, H_k) + \sum_{\alpha=1}^{(N^2-N)/2} (W^*{}^{\alpha} (H_r, \tilde{E}_{\alpha}) + W^{\alpha} (H_r, \tilde{E}_{-\alpha}))  
 \nonumber\\
 =&  V_r  ,
\end{align}
since 
$(H_j, \tilde{E}_{\alpha})={\rm tr}(H_j \tilde{E}_{\alpha})=0$. 

For any Lie algebra valued function $\mathscr{V}_N(x)$, we obtain the identity:
\begin{align}
  \mathscr{V}_N 
=& \sum_{A=1}^{N^2-1} \mathscr{V}_A T_A
\nonumber\\
=&  \tilde{\mathscr{V}_N} + (\mathscr{V}_N,H_r) H_r 
+  \frac{1}{\alpha_r^2}  [H_r , [H_r , \mathscr{V}_N]]
\nonumber\\
=&  \tilde{\mathscr{V}_N} + (\mathscr{V}_N,H_r) H_r 
+  \frac{2(N-1)}{N}  [H_r , [H_r , \mathscr{V}_N]]
 .
\label{idv4}
\end{align}

\section{CFNS decomposition}\label{section:CFNS}

This section is a summary of the results obtained in \cite{KSM08}, which is added just for the convenience of readers.

\subsection{Maximal case}

In the maximal case, we introduce a full set of color fields $\bm n_j(x)$ ($j=1, \cdots, r=N-1$) according to the adjoint orbit representation:
\begin{align}
  \bm{n}_j(x) = U^{\dagger}(x) H_j U(x) , \quad j \in \{ 1, 2, \cdots, r \} 
 ,
\label{ador}
\end{align}  
where  $r=\text{rank} SU(N)=N-1$ and $H_j$ are Cartan subalgebra.
The fields $\bm{n}_j(x)$ defined in this way are  unit vectors, since
\begin{align}
  (\bm{n}_j(x),\bm{n}_k(x)) &= 2 {\rm tr}(\bm{n}_j(x)\bm{n}_k(x))
=  2 {\rm tr}(U^{\dagger}(x) H_j U(x) U^{\dagger}(x) H_k U(x)) 
\nonumber\\&
=  2 {\rm tr}( H_j  H_k ) 
= (H_j ,H_k) = \delta_{jk}
 .
\end{align}
These unit vectors mutually commute, 
\begin{align}
  [ \bm{n}_j(x), \bm{n}_k(x)] = 0, \quad j,k \in \{ 1, 2, \cdots, r \} 
 ,
\label{com}
\end{align}
since $H_j$ are Cartan subalgebra obeying 
\begin{align}
  [ H_j ,H_k ] = 0, \quad j,k \in \{ 1, 2, \cdots, r \} 
 .
\end{align}

Once such a set of color  fields $\bm n_j(x)$ is given, the original gauge field  has the decomposition:
\begin{align}
\mathscr A_\mu(x)
 =\mathscr V_\mu(x)
  +\mathscr X_\mu(x)
,
\end{align}
where the respective components $\mathscr V_\mu(x)$ and $\mathscr X_\mu(x)$ are specified by two defining equations (conditions):

(I) all $\bm{n}_j(x)$ are covariant constant in the background $\mathscr{V}_\mu(x)$:
\begin{align}
  0 = D_\mu[\mathscr{V}] \bm{n}_j(x) 
:=\partial_\mu \bm{n}_j(x) -  ig [\mathscr{V}_\mu(x), \bm{n}_j(x)]
\quad (j=1,2, \cdots, r) 
 ,
\label{defVL}
\end{align}

(II)  $\mathscr{X}_\mu(x)$  is orthogonal to all $\bm{n}_j(x)$:
\begin{align}
  (\mathscr{X}_\mu(x), \bm{n}_j(x)) 
 := 2{\rm tr}(\mathscr{X}_\mu(x) \bm{n}_j(x) )  = \mathscr{X}_\mu^A(x) n_j^A(x)  = 0  
\quad (j=1,2, \cdots, r) 
\label{defXL}
 . 
\end{align}

First, we determine the $\mathscr{X}_\mu$ field by solving the defining equations. 
We apply the identity (\ref{idv}) to $\mathscr{X}_\mu$ and use the second defining equation (\ref{defXL}) to obtain   
\begin{align}
 \mathscr{X}_\mu = \sum_{j=1}^{r} (\mathscr{X}_\mu,\bm{n}_j)\bm{n}_j +   \sum_{j=1}^{r}  [\bm{n}_j, [\bm{n}_j, \mathscr{X}_\mu]]
=  \sum_{j=1}^{r}   [\bm{n}_j, [\bm{n}_j, \mathscr{X}_\mu]]
 .
\end{align}
Then we take into account the first defining equation: 
\begin{align}
 \mathscr{D}_\mu[\mathscr{A}]\bm{n}_j 
 =& \partial_\mu \bm{n}_j - ig[\mathscr{A}_\mu, \bm{n}_j]
\nonumber\\
=& \mathscr{D}_\mu[\mathscr{V}]\bm{n}_j - ig [\mathscr{X}_\mu, \bm{n}_j] 
\nonumber\\ 
=& - ig [\mathscr{X}_\mu, \bm{n}_j] =  ig [ \bm{n}_j, \mathscr{X}_\mu] 
 .
\end{align}
Thus $\mathscr{X}_\mu(x)$ is expressed in terms of $\mathscr{A}_\mu(x)$ and $\bm{n}_j(x)$ as
\begin{align}
 \mathscr{X}_\mu(x) = -ig^{-1} \sum_{j=1}^{r}  [\bm{n}_j(x), \mathscr{D}_\mu[\mathscr{A}]\bm{n}_j(x) ]
 .
\end{align}

Next, the $\mathscr{V}_\mu$ field is expressed in terms of $\mathscr{A}_\mu(x)$ and $\bm{n}_j(x)$: 
\begin{align}
  \mathscr{V}_\mu(x) 
  =& \mathscr{A}_\mu(x)  - \mathscr{X}_\mu(x)  
  \nonumber\\
  =& \mathscr{A}_\mu(x)  + ig^{-1} \sum_{j=1}^{r}  [\bm{n}_j(x), \mathscr{D}_\mu[\mathscr{A}]\bm{n}_j(x) ] 
  \nonumber\\
  =& \mathscr{A}_\mu(x) - \sum_{j=1}^{r}  [\bm{n}_j(x), [ \bm{n}_j(x), \mathscr{A}_\mu(x) ] ]
+ ig^{-1} \sum_{j=1}^{r}    [\bm{n}_j(x) , \partial_\mu  \bm{n}_j(x) ]
 .
\label{defV11}
\end{align}
Now we apply the identity (\ref{idv}) to $\mathscr{A}_\mu$ to obtain 
\begin{align}
\mathscr{V}_\mu(x)
=\sum_{j=1}^{r}(\mathscr{A}_\mu(x),\bm{n}_j(x))
\bm{n}_j(x)+ig^{-1} \sum_{j=1}^{r} [\bm{n}_j(x) , \partial_\mu  \bm{n}_j(x)].
\label{defV}
\end{align}
Thus, $\mathscr{V}_\mu(x)$ and $\mathscr{X}_\mu(x)$ are  written in terms of $\mathscr A_\mu(x)$, once $\bm n_j(x)$ is given as a functional of $\mathscr A_\mu(x)$.

It should be remarked that the background field $\mathscr{V}_\mu(x)$ contains a part $\mathscr{C}_\mu(x)$ which commutes with all $\bm{n}_j(x)$:
\begin{align}
    [ \mathscr{C}_\mu(x), \bm{n}_j(x) ] = 0 
    \quad (j=1,2, \cdots, r=N-1) 
\label{specC}
 .
\end{align}
Such a commutative part (or a parallel part in the vector form) $\mathscr{C}_\mu(x)$  in  $\mathscr{V}_\mu(x)$ is not determined uniquely from the first defining equation (\ref{defVL}) alone. But it was determined by the second defining equation as shown above. 
In view of this,  we further decompose $\mathscr V_\mu(x)$ into $\mathscr C_\mu(x)$ and $\mathscr B_\mu(x)$:
\begin{align}
 \mathscr V_\mu(x)
 =\mathscr C_\mu(x)
  +\mathscr B_\mu(x)
 .
\end{align}
Applying the identity (\ref{idv}) to $\mathscr{C}_\mu(x)$ and by taking into account (\ref{specC}), we obtain
\begin{align}
 \mathscr{C}_\mu(x)
 = \sum_{j=1}^{r}  (\mathscr{C}_\mu(x),\bm{n}_j(x)) \bm{n}_j(x) 
 . 
\end{align}
If the remaining part $\mathscr{B}_\mu(x)$ which is not commutative $[\mathscr{B}_\mu(x),\bm{n}_j(x)]  \ne 0$ is perpendicular to all $\bm{n}_j(x)$:
\begin{align}
 (\mathscr{B}_\mu(x),  \bm{n}_j(x))
= 2{\rm tr}(\mathscr{B}_\mu(x) \bm{n}_j(x) )=0 
\quad (j=1,2, \cdots, r)
\label{defBL}
 ,
\end{align}
then we have   
\begin{align}
  (\mathscr{A}_\mu(x),\bm{n}_j(x)) 
=  (\mathscr{V}_\mu(x),\bm{n}_j(x)) 
= (\mathscr{C}_\mu(x),\bm{n}_j(x)) 
 . 
\end{align}
Consequently, the parallel part $\mathscr{C}_\mu(x)$ reads
\begin{align}
 \mathscr{C}_\mu(x)
 = \sum_{j=1}^{r}  (\mathscr{A}_\mu(x),\bm{n}_j(x)) \bm{n}_j(x)
 . 
\end{align}
and the perpendicular part $\mathscr{B}_\mu(x)$ is determined as
\footnote{
The SU(2) version in the vector form reads 
$\mathbb{B}_\mu(x)
=g^{-1}  \partial_\mu  \bm{n}(x) \times  \bm{n}(x)$.
} 
\begin{align}
  \mathscr{B}_\mu(x) 
  =   ig^{-1}  \sum_{j=1}^{r} [ \bm{n}_j(x), \partial_\mu  \bm{n}_j(x) ]  
 .
\label{B}
\end{align}
In fact, it is easy to check that this expression indeed satisfies (\ref{defBL}) and  
\begin{align}
 D_\mu[\mathscr{B}] \bm{n}_j(x) 
=\partial_\mu \bm{n}_j(x) -  ig [\mathscr{B}_\mu(x), \bm{n}_j(x)] = 0
\quad (j=1,2, \cdots, r) 
 ,
\label{defBL2}
\end{align}

Thus, once full set of color fields $\bm n_j(x)$ is given, the original gauge field  has the decomposition in the Lie algebra form: 
\begin{subequations}
\begin{align}
\mathscr A_\mu(x)
 =\mathscr V_\mu(x)
  +\mathscr X_\mu(x)
 =\mathscr C_\mu(x)
  +\mathscr B_\mu(x)
  +\mathscr X_\mu(x)
,
\end{align}
where each part is expressed  in terms of $\mathscr{A}_\mu(x)$ and $\bm{n}_j(x)$ as
\begin{align}
  \mathscr{C}_\mu(x) =&   \sum_{j=1}^{N-1}( \mathscr{A}_\mu(x),\bm{n}_j(x))  \bm{n}_j(x) 
= \sum_{j=1}^{N-1}  c_\mu^j(x) \bm{n}_j(x) ,
\\
  \mathscr{B}_\mu(x) =&   
   ig^{-1} \sum_{j=1}^{N-1} [\bm{n}_j(x), \partial_\mu  \bm{n}_j(x)] ,
\\
 \mathscr{X}_\mu(x) =& -ig^{-1}   \sum_{j=1}^{N-1}  [\bm{n}_j(x), \mathscr{D}_\mu[\mathscr{A}]\bm{n}_j(x) ]
 .
\end{align}
\end{subequations}
In what follows, the summation over the index $j$ should be understood when it is repeated, unless otherwise stated.

\subsection{Minimal case}

Now we consider the minimal case for $SU(N)$. 
In this case,  $\mathscr A_\mu$ is decomposed as
\begin{align}
 \mathscr{A}_\mu(x) = \mathscr{V}_\mu(x) + \mathscr{X}_\mu(x) 
,
\end{align}
using  only a single color  field $\bm n_r (r=N-1)$: 
\begin{align}
 \bm h(x) :=\bm n_r(x) =U^\dagger(x) H_r U(x)
  ,
\end{align}
where $H_r$ is the last diagonal matrix.

The respective components $\mathscr V_\mu(x)$ and $\mathscr X_\mu(x)$ are specified by two defining equations (conditions):

\noindent
(I)  $\bm{h}(x)$ is covariant constant in the background $\mathscr{V}_\mu(x)$:
\begin{align}
  0 = D_\mu[\mathscr{V}] \bm{h}(x) 
:=\partial_\mu \bm{h}(x) -  ig [\mathscr{V}_\mu(x), \bm{h}(x)]
 ,
\label{defVL2}
\end{align}
(II)  $\mathscr{X}_\mu(x)$  is orthogonal to $\bm{h}(x)$:
\begin{align}
  (\mathscr{X}_\mu(x), \bm{h}(x)) 
 := 2{\rm tr}(\mathscr{X}_\mu(x) \bm{h}(x) ) 
 = \mathscr{X}_\mu^A(x) \bm{h}^A(x)  = 0  
\label{defXL2}
 . 
\end{align}
(II) $\mathscr{X}_\mu(x)$ does not have a part $\tilde{\mathscr{X}}_\mu(x)$ which commutes with $\bm{h}(x)$:
\begin{align}
  \tilde{\mathscr{X}}_\mu(x) = 0
\label{defXL3}
 . 
\end{align}

First, we apply the identity (\ref{idv2}) to $\mathscr{X}_\mu(x)$ and use the second and third defining equations to obtain
\begin{align}
  \mathscr{X}_\mu(x) 
=&  \tilde{\mathscr{X}}_\mu(x)  + (\mathscr{X}_\mu(x),\bm{h})  \bm{h}
+   \frac{2(N-1)}{N}  [\bm{h} , [\bm{h} , \mathscr{X}_\mu(x)]]
\nonumber\\
=&    \frac{2(N-1)}{N}  [\bm{h} , [\bm{h} , \mathscr{X}_\mu(x)]]
 .
\end{align}

Then we take into account the first defining equation: 
\begin{align}
 \mathscr{D}_\mu[\mathscr{A}]\bm{h}  
 =& \partial_\mu \bm{h}  - ig[\mathscr{A}_\mu, \bm{h} ]
\nonumber\\
=& \mathscr{D}_\mu[\mathscr{V}]\bm{h}  - ig [\mathscr{X}_\mu, \bm{h} ] 
\nonumber\\ 
=& - ig [\mathscr{X}_\mu, \bm{h} ] =  ig [ \bm{h} , \mathscr{X}_\mu] 
 .
\end{align}
Thus $\mathscr{X}_\mu(x)$ is expressed in terms of $\mathscr{A}_\mu(x)$ and $\bm{h}(x)$ as
\begin{align}
 \mathscr{X}_\mu(x) = -ig^{-1}  \frac{2(N-1)}{N}  [\bm{h}(x), \mathscr{D}_\mu[\mathscr{A}]\bm{h}(x) ]
 .
\end{align}

Next, the $\mathscr{V}_\mu$ field is expressed in terms of $\mathscr{A}_\mu(x)$ and $\bm{h}(x)$: 
\begin{align}
  \mathscr{V}_\mu(x) 
  =& \mathscr{A}_\mu(x)  - \mathscr{X}_\mu(x)  
  \nonumber\\
  =& \mathscr{A}_\mu(x)  + ig^{-1}  \frac{2(N-1)}{N}  [\bm{h}(x), \mathscr{D}_\mu[\mathscr{A}]\bm{h}(x) ]
  \nonumber\\
  =& \mathscr{A}_\mu(x) - \frac{2(N-1)}{N}   [\bm{h}(x), [ \bm{h}(x), \mathscr{A}_\mu(x) ] ]
  \nonumber\\
  &+ ig^{-1} \frac{2(N-1)}{N}   [\bm{h}(x) , \partial_\mu  \bm{h}(x) ]
 .
\label{defV12}
\end{align}
Thus, $\mathscr{V}_\mu(x)$ and $\mathscr{X}_\mu(x)$ are  written in terms of $\mathscr A_\mu(x)$, once $\bm h(x)$ is given as a functional of $\mathscr A_\mu(x)$.

Now we apply the identity (\ref{idv2}) to $\mathscr{A}_\mu$ to obtain 
\begin{align}
\mathscr{V}_\mu(x)
= \tilde{\mathscr{A}}_\mu(x) + (\mathscr{A}_\mu(x),\bm{h}(x))
\bm{h}(x)
+ig^{-1} \frac{2(N-1)}{N}   [\bm{h}(x) , \partial_\mu  \bm{h}(x) ]
 .
\label{defV2}
\end{align}

We further decompose $\mathscr V_\mu(x)$ into $\mathscr C_\mu(x)$ and $\mathscr B_\mu(x)$:
\begin{align}
 \mathscr V_\mu(x)
 =\mathscr C_\mu(x)
  +\mathscr B_\mu(x)
 .
\end{align}
where  $\mathscr{C}_\mu(x)$ commutes with $\bm{h}_(x)$:
\begin{align}
    [ \mathscr{C}_\mu(x), \bm{h}(x) ] = 0 
    \quad (j=1,2, \cdots, r=N-1) 
\label{specC2}
 ,
\end{align}
and the remaining part $\mathscr{B}_\mu(x)$ which is not commutative $[\mathscr{B}_\mu(x),\bm{n}_j(x)]  \ne 0$ is perpendicular to  $\bm{h}(x)$:
\begin{align}
 (\mathscr{B}_\mu(x),  \bm{h}(x))
= 2{\rm tr}(\mathscr{B}_\mu(x) \bm{h}(x) )=0 
\label{defBL22}
 ,
\end{align}
which leads to    
\begin{align}
  (\mathscr{A}_\mu(x),\bm{h}(x)) 
=  (\mathscr{V}_\mu(x),\bm{h}(x)) 
= (\mathscr{C}_\mu(x),\bm{h}(x)) 
 . 
\end{align}
Thus, once a single color field $\bm{h}(x)$ is given, we have the decomposition:
\begin{subequations}
\begin{align}
\mathscr A_\mu(x)
 =& \mathscr V_\mu(x)
  +\mathscr X_\mu(x)
 =\mathscr C_\mu(x)
  +\mathscr B_\mu(x)
  +\mathscr X_\mu(x)
 ,
\\
  \mathscr{C}_\mu(x)
=& \mathscr{A}_\mu(x) - \frac{2(N-1)}{N}   [\bm{h}(x), [ \bm{h}(x), \mathscr{A}_\mu(x) ] ]
,
\\
 \mathscr{B}_\mu(x)
=& ig^{-1} \frac{2(N-1)}{N}[\bm{h}(x) , \partial_\mu  \bm{h}(x) ]
,
\\
 \mathscr{X}_\mu(x) =& -ig^{-1}  \frac{2(N-1)}{N}  [\bm{h}(x), \mathscr{D}_\mu[\mathscr{A}]\bm{h}(x) ]
 .
\end{align}
\end{subequations}

\newpage
\section{Field strength}\label{section:field-strength}

This section is a summary of the results obtained in \cite{KSM08}, which is added just for the convenience of readers.

\subsection{Maximal case}

The field strength is rewritten in terms of new variables as 
\begin{align}
 & \mathscr{F}_{\mu\nu}[\mathscr{A}] 
\nonumber\\
:=& \partial_\mu \mathscr{A}_\nu - \partial_\nu \mathscr{A}_\mu - ig [\mathscr{A}_\mu , \mathscr{A}_\nu ]
\nonumber\\
=& \mathscr{F}_{\mu\nu}[\mathscr{V}]
+ \partial_\mu  \mathscr{X}_\nu  - \partial_\nu  \mathscr{X}_\mu 
- ig [ \mathscr{V}_\mu ,   \mathscr{X}_\nu ]
- ig [  \mathscr{X}_\mu ,  \mathscr{V}_\nu  ]
 - ig [  \mathscr{X}_\mu ,   \mathscr{X}_\nu ]
\nonumber\\
=& \mathscr{F}_{\mu\nu}[\mathscr{V}]
+ D_\mu[\mathscr{V}]  \mathscr{X}_\nu  - D_\nu[\mathscr{V}]  \mathscr{X}_\mu  - ig [  \mathscr{X}_\mu ,   \mathscr{X}_\nu ]
 ,
\end{align}
Here $\mathscr{F}_{\mu\nu}[\mathscr{V}]$ is further decomposed as (omitting the summation symbol over $j$):
\begin{align}
 &  \mathscr{F}_{\mu\nu}[\mathscr{V}]
\nonumber\\
 :=& \partial_\mu \mathscr{V}_\nu - \partial_\nu \mathscr{V}_\mu - ig [\mathscr{V}_\mu , \mathscr{V}_\nu ]
\nonumber\\
=& \mathscr{F}_{\mu\nu}[\mathscr{B}] 
+ \partial_\mu  \mathscr{C}_\nu  - \partial_\nu  \mathscr{C}_\mu  
- ig [ \mathscr{B}_\mu  ,  \mathscr{C}_\nu  ]
- ig [ \mathscr{C}_\mu  ,  \mathscr{B}_\nu  ]
- ig [ \mathscr{C}_\mu  ,  \mathscr{C}_\nu  ]
\nonumber\\  
=& \mathscr{F}_{\mu\nu}[\mathscr{B}] 
+ \bm{n}_j \partial_\mu  c_\nu^j  - \bm{n}_j \partial_\nu  c_\mu^j  
\nonumber\\  
&+  c_\nu^j \partial_\mu  \bm{n}_j  
 - ig [ \mathscr{B}_\mu  ,  c_\nu^j \bm{n}_j ]
- c_\mu^j \partial_\nu  \bm{n}_j
+ ig [\mathscr{B}_\nu , c_\mu^j \bm{n}_j  ]
- ig [ c_\mu^j \bm{n}_j  ,  c_\nu^k \bm{n}_k  ]
\nonumber\\  
=& \mathscr{F}_{\mu\nu}[\mathscr{B}] 
+ \bm{n}_j \partial_\mu  c_\nu^j  - \bm{n}_j \partial_\nu  c_\mu^j  
\nonumber\\  
:=& \mathscr{H}_{\mu\nu} + \mathscr{E}_{\mu\nu} 
 .
\end{align}
Here  we have used (\ref{com}) 
and (\ref{defBL2}) in the last step in simplifying $\mathscr{E}_{\mu\nu}$.
Therefore we obtain the decomposition of the field strength:
\begin{subequations}
\begin{align}
  \mathscr{F}_{\mu\nu}[\mathscr{A}]  
=  \mathscr{E}_{\mu\nu} + \mathscr{H}_{\mu\nu} + D_\mu[\mathscr{V}] \mathscr{X}_\nu - D_\nu[\mathscr{V}] \mathscr{X}_\mu - ig [\mathscr{X}_\mu , \mathscr{X}_\nu ], 
\end{align}
where we have defined
\begin{align}
  \mathscr{E}_{\mu\nu} :=& \sum_{j=1}^{N-1} \bm{n}_j E_{\mu\nu}^j , \quad
E_{\mu\nu}^j := \partial_\mu c_\nu^j - \partial_\nu c_\mu^j ,
\\
  \mathscr{H}_{\mu\nu} :=& \mathscr{F}_{\mu\nu}[\mathscr{B}]
 =  \partial_\mu \mathscr{B}_\nu - \partial_\nu \mathscr{B}_\mu
  -i g [\mathscr{B}_\mu , \mathscr{B}_\nu] .
\end{align}
\end{subequations}
Here $\mathscr{H}_{\mu\nu}$ is simplified for
$
\mathscr{B}_\mu(x) =    
   ig^{-1} \sum_{j=1}^{N-1} [\bm{n}_j(x), \partial_\mu  \bm{n}_j(x)]  
$
as (omitting the summation symbol over $j$): 
\begin{subequations}
\begin{align}
 & \partial_\mu \mathscr{B}_\nu - \partial_\nu \mathscr{B}_\mu
\nonumber\\
=&  ig^{-1}  [ \partial_\mu \bm{n}_j , \partial_\nu \bm{n}_j] - (\mu \leftrightarrow \nu) 
\label{a}
\\
=&  ig^{-1}  (ig)^2 [ [\mathscr{B}_\mu ,\bm{n}_j], [\mathscr{B}_\nu ,\bm{n}_j]] - (\mu \leftrightarrow \nu) 
\label{b}
\\
=&  ig    [[ \bm{n}_j, [\mathscr{B}_\nu ,\bm{n}_j]], \mathscr{B}_\mu]
+ [[[ \mathscr{B}_\nu ,\bm{n}_j], \mathscr{B}_\mu], \bm{n}_j]  
- (\mu \leftrightarrow \nu)  
\label{c}
\\
=& 
   ig  [ \mathscr{B}_\mu, [\bm{n}_j, [ \bm{n}_j, \mathscr{B}_\nu ]]]  
+   [\bm{n}_j, [[ \bm{n}_j, \mathscr{B}_\nu ], \mathscr{B}_\mu]] - (\mu \leftrightarrow \nu) 
\label{d}
\\
=& 
   ig  [ \mathscr{B}_\mu, [\bm{n}_j, [ \bm{n}_j, \mathscr{B}_\nu ]]] - (\mu \leftrightarrow \nu) 
- ig    [\bm{n}_j, [\bm{n}_j, [\mathscr{B}_\mu  , \mathscr{B}_\nu]]] 
\label{e}
\\
=& 
   ig   ([ \mathscr{B}_\mu,  \mathscr{B}_\nu ]  - [ \mathscr{B}_\nu,  \mathscr{B}_\mu ])
- ig    [\bm{n}_j, [\bm{n}_j, [\mathscr{B}_\mu  , \mathscr{B}_\nu]]] 
\label{f}
\\
=& 
  2ig   [ \mathscr{B}_\mu,  \mathscr{B}_\nu ]  
- ig  [[\mathscr{B}_\mu  , \mathscr{B}_\nu]
+   \bm{n}_j (\bm{n}_j, ig [\mathscr{B}_\mu  , \mathscr{B}_\nu]) 
\label{g}
\\
=&  i g [\mathscr{B}_\mu , \mathscr{B}_\nu] 
+ig   \bm{n}_j(\bm{n}_j,[\mathscr{B}_\mu, \mathscr{B}_\nu])
,
\label{h}
\end{align}
\end{subequations}
where we have used 
(\ref{B}) in (\ref{a}), 
(\ref{defBL2}) in (\ref{b}), 
the Jacobi identity for $\mathscr{B}_\mu$, $\bm{n}_j$ and $[\mathscr{B}_\nu ,\bm{n}_j]$ in (\ref{c}), 
interchanged the commutator in (\ref{d}), 
the Jacobi identity in (\ref{e}), 
the algebraic identity (\ref{idv}) with (\ref{defBL}) to obtain the first term of (\ref{f})  
and again 
the algebraic identity (\ref{idv}) in (\ref{g}).
Therefore we obtain 
\begin{align}
 \mathscr{H}_{\mu\nu}  
:= \partial_\mu \mathscr{B}_\nu - \partial_\nu \mathscr{B}_\mu
  -i g [\mathscr{B}_\mu , \mathscr{B}_\nu] 
=  ig   \bm{n}_j(\bm{n}_j,[\mathscr{B}_\mu, \mathscr{B}_\nu])
 .
\end{align}
Thus we find $\mathscr{H}_{\mu\nu}$ is written as the linear combination of all the color fields just like $\mathscr{E}_{\mu\nu}$:
\begin{align}
 \mathscr{H}_{\mu\nu} = \mathscr{F}_{\mu\nu}[\mathscr{B}]  = \sum_{j=1}^{N-1} \bm{n}_j H^j_{\mu\nu}
, 
\quad 
H^j_{\mu\nu} = ig  (\bm{n}_j,[\mathscr{B}_\mu, \mathscr{B}_\nu]) .
\end{align}

\subsection{Minimal case}

The field strength $\mathscr{F}_{\mu\nu}[\mathscr{V}]$ reads
\begin{align}
 & \mathscr{F}_{\mu\nu}[\mathscr{V}] 
\nonumber\\
:=& \partial_\mu \mathscr{V}_\nu - \partial_\nu \mathscr{V}_\mu - ig [\mathscr{V}_\mu , \mathscr{V}_\nu ]
\nonumber\\
=& \mathscr{F}_{\mu\nu}[\mathscr{B}]
+ \partial_\mu  \mathscr{C}_\nu  - \partial_\nu  \mathscr{C}_\mu 
- ig [ \mathscr{B}_\mu ,   \mathscr{C}_\nu ]
- ig [  \mathscr{C}_\mu ,  \mathscr{B}_\nu  ]
 - ig [  \mathscr{C}_\mu ,   \mathscr{C}_\nu ]
\nonumber\\
=& \mathscr{F}_{\mu\nu}[\mathscr{B}]
+ D_\mu[\mathscr{B}]  \mathscr{C}_\nu  - D_\nu[\mathscr{B}]  \mathscr{C}_\mu  - ig [  \mathscr{C}_\mu ,   \mathscr{C}_\nu ]
 ,
\end{align}
where 
\begin{align}
 \mathscr{F}_{\mu\nu}[\mathscr{B}]=\mathscr{H}_{\mu\nu} := \partial_\mu \mathscr{B}_\nu - \partial_\nu \mathscr{B}_\mu  -i g [\mathscr{B}_\mu , \mathscr{B}_\nu] 
 .
\end{align}

We focus on the parallel part of the field strength $\mathscr{F}_{\mu\nu}[\mathscr{V}]$:
\begin{align}
& {\rm tr}( \bm{h}\mathscr{F}_{\mu\nu}[\mathscr{V}] )
\nonumber\\
=& {\rm tr}( \bm{h}\mathscr{F}_{\mu\nu}[\mathscr{B}] )
+ {\rm tr}( \bm{h} D_\mu[\mathscr{B}]  \mathscr{C}_\nu)
- {\rm tr}( \bm{h} D_\nu[\mathscr{B}]  \mathscr{C}_\mu)
- ig {\rm tr}( \bm{h} [  \mathscr{C}_\mu ,   \mathscr{C}_\nu ])
\nonumber\\
=& {\rm tr}( \bm{h}\mathscr{F}_{\mu\nu}[\mathscr{B}] )
- {\rm tr}( (D_\mu[\mathscr{B}] \bm{h})  \mathscr{C}_\nu)
+ \partial_\mu {\rm tr}(  \bm{h} \mathscr{C}_\nu)
\nonumber\\&
+ {\rm tr}( (D_\nu[\mathscr{B}] \bm{h})   \mathscr{C}_\mu)
-  \partial_\mu  {\rm tr} ( \bm{h} \mathscr{C}_\mu)
-  ig {\rm tr}( \mathscr{C}_\nu  [ \bm{h},  \mathscr{C}_\mu ])
\nonumber\\
=& {\rm tr}( \bm{h}\mathscr{F}_{\mu\nu}[\mathscr{B}] )
+ \partial_\mu {\rm tr}(  \bm{h} \mathscr{C}_\nu)
-  \partial_\mu  {\rm tr} ( \bm{h} \mathscr{C}_\mu)
 ,
\end{align}
where we have used the defining equations for $\mathscr{B}_\mu$ and $\mathscr{C}_\mu$:
$D_\mu[\mathscr{B}] \bm{h}=0$ and $[ \bm{h},  \mathscr{C}_\mu ]=0$.

Then we focus on the parallel part of the field strength $\mathscr{F}_{\mu\nu}[\mathscr{B}]=\mathscr{H}_{\mu\nu}$:
\begin{subequations}
\begin{align}
 {\rm tr}( \bm{h} \mathscr{H}_{\mu\nu})
 =& {\rm tr}( \bm{h}\mathscr{F}_{\mu\nu}[\mathscr{B}] )
 \nonumber\\
=& 
 {\rm tr}( \bm{h} \partial_\mu \mathscr{B}_\nu - \bm{h} \partial_\nu \mathscr{B}_\mu
  -i g \bm{h} [\mathscr{B}_\mu , \mathscr{B}_\nu])
\\
=& 
 {\rm tr}( \bm{h} \partial_\mu \mathscr{B}_\nu - \bm{h} \partial_\nu \mathscr{B}_\mu
  +i g \mathscr{B}_\nu  [\mathscr{B}_\mu , \bm{h}])
\\
=& 
 {\rm tr}( \bm{h} \partial_\mu \mathscr{B}_\nu - \bm{h} \partial_\nu \mathscr{B}_\mu
  + \mathscr{B}_\nu \partial_\mu \bm{h} )
\\
=& 
 \partial_\mu {\rm tr}(\bm{h} \mathscr{B}_\nu) - {\rm tr}(\bm{h} \partial_\nu \mathscr{B}_\mu )
\\
=& 
  {\rm tr}(\mathscr{B}_\mu  \partial_\nu  \bm{h}) - \partial_\nu {\rm tr}(\bm{h} \mathscr{B}_\mu) 
\\
=& 
  {\rm tr}(\mathscr{B}_\mu  \partial_\nu  \bm{h}) 
 ,
\end{align}
\end{subequations}
where we have used ${\rm tr}(\bm{h} \mathscr{B}_\mu)=0$ twice. 

On the other hand, we find
\begin{subequations}
\begin{align}
 {\rm tr}( \bm{h} [\partial_\mu \bm{h}, \partial_\nu \bm{h}])
=& 
 {\rm tr}( \bm{h} \partial_\mu \bm{h} \partial_\nu \bm{h} - \bm{h} \partial_\nu \bm{h} \partial_\mu \bm{h} )
\\
=& 
 {\rm tr}( (\bm{h} \partial_\mu \bm{h}  -  \partial_\mu \bm{h}  \bm{h} ) \partial_\nu \bm{h} )
\\
=& 
 {\rm tr}( [ \bm{h}, \partial_\mu \bm{h}] \partial_\nu \bm{h} )
 .
\end{align}
\end{subequations}
By using the explicit form of 
\begin{align}
 \mathscr{B}_\mu 
 = ig^{-1} \frac{2(N-1)}{N} [ \bm{h}, \partial_\mu \bm{h}] 
 ,
\end{align}
therefore, both expressions are connected as
\begin{align}
 {\rm tr}( \bm{h} \mathscr{H}_{\mu\nu})
 =&   {\rm tr}( \bm{h}\mathscr{F}_{\mu\nu}[\mathscr{B}] )
 \nonumber\\
=&   \frac{2(N-1)}{N} {\rm tr}( ig^{-1}[ \bm{h}, \partial_\mu \bm{h}] \partial_\nu \bm{h} )
 \nonumber\\
=&   \frac{2(N-1)}{N} 
 {\rm tr}(ig^{-1} \bm{h} [\partial_\mu \bm{h}, \partial_\nu \bm{h}])
 .
\end{align}
Note that  
$\bm{h}\mathscr{F}_{\mu\nu}[\mathscr{B}] \ne \frac{2(N-1)}{N} 
 ig^{-1} \bm{h} [\partial_\mu \bm{h}, \partial_\nu \bm{h}] $.

The same result is obtained by calculating explicitly the field strength
$
\mathscr{F}_{\mu\nu}[\mathscr{B}]  
  :=  \partial_\mu \mathscr{B}_\nu - \partial_\nu \mathscr{B}_\mu  -i g [\mathscr{B}_\mu , \mathscr{B}_\nu] 
$
using the property of $\bm{h}$: 
\begin{align}
 \bm{h}\bm{h} = \frac{1}{2N} {\bf 1} + \frac{2-N}{\sqrt{2N(N-1)}} \bm{h} 
 ,
\end{align}
which follows from the relation for the last Cartan generator:
\begin{align}
 H_{N-1} H_{N-1}  = \frac{1}{2N} {\bf 1} + \frac{2-N}{\sqrt{2N(N-1)}} H_{N-1}
 .
\end{align}

For a fundamental representation, a weight vector is given by
\begin{align}
\bm{\Lambda} = \nu_N \equiv \left( 0, 0, \cdots, 0, \frac{1-N}{\sqrt{2N(N-1)}} \right) 
 .
\end{align}
Using 
\begin{align}
 H_{N-1} =  \frac{1}{\sqrt{2N(N-1)}} {\rm diag}(1, 1, \cdots, 1, -N+1) 
 ,
\end{align}
we have
\begin{align}
 \mathcal{H} 
=& \sum_{j=1}^{N-1} \Lambda_{j} H_{j} 
= \sum_{j=1}^{N-1}  (\nu_N)_j  H_{j} 
\nonumber\\
=& (\nu_N)_{N-1}   H_{N-1} 
=  \frac{1}{2}  {\rm diag}(\frac{-1}{N}, \frac{-1}{N}, \cdots, \frac{-1}{N}, \frac{N-1}{N}) 
 .
\end{align}

The $\bm{m}$ field in the non-Abelian Stokes theorem reads
\begin{align}
 \bm{m}(x) = (\nu_N)_{N-1} \bm{h}(x)
=  \frac{1-N}{\sqrt{2N(N-1)}} \bm{h}(x) 
 .
\end{align}
By using the explicit form of 
\begin{align}
 \mathscr{B}_\mu 
 = 4 ig^{-1} [ \bm{m}, \partial_\mu \bm{m}] 
 ,
\end{align}
we have
\begin{align}
 {\rm tr}( \bm{m} \mathscr{H}_{\mu\nu})
 =&   {\rm tr}( \bm{m}\mathscr{F}_{\mu\nu}[\mathscr{B}] )
 \nonumber\\
=&   4 {\rm tr}( ig^{-1}[ \bm{m}, \partial_\mu \bm{m}] \partial_\nu \bm{m} )
 \nonumber\\
=&   4
 {\rm tr}(ig^{-1} \bm{m} [\partial_\mu \bm{m}, \partial_\nu \bm{m}])
 .
 \label{mF}
\end{align}


\end{document}